\documentclass[%
reprint,
 amsmath,amssymb,
 aps,
]{revtex4-1}

\usepackage{color}
\usepackage{graphicx}% Include figure files
\usepackage{dcolumn}% Align table columns on decimal point
\usepackage{bm}% bold math
\usepackage{hyperref}% add hypertext capabilities
\def\Vec#1{\boldsymbol #1}

\begin{document}

%\preprint{APS/123-QED}

\title{
Berezinskii-Kosterlitz-Thouless Transition of Spin-1 Spinor Bose Gases in the Presence of the Quadratic Zeeman Effect
}

\author{
Michikazu Kobayashi$^1$
}
\affiliation{%
$^1$Department of Physics, Kyoto University, Oiwake-cho, Kitashirakawa, Sakyo-ku, Kyoto 606-8502, Japan
}%

\date{\today}% It is always \today, today,
             %  but any date may be explicitly specified

\begin{abstract}
We numerically study the Berezinskii-Kosterlitz-Thouless (BKT) transition of a spin-1 spinor Bose gas under the quadratic Zeeman effect.
A calculation of the mass and spin superfluid densities shows that (i) the BKT transition occurs only when vortices are classified by the integer group $\mathbb{Z}$, and $\mathbb{Z}_2$ vortices do not contribute to the BKT transition, (ii) the two BKT transition temperatures for mass and spin superfluid densities are different for a positive quadratic Zeeman effect and equal for a negative quadratic Zeeman effect, and (iii) the universal relation of the superfluid densities at the BKT transition temperature is changed when multiple kinds of vortices contribute to the transition.
We have further found that (iv) spin-singlet pairs in non-magnetic states show the quasi-off-diagonal-long-range order at the different temperature lower than the BKT transition temperature, giving the new universal relation of the superfluid density.
\end{abstract}


\maketitle

\section{Introduction}

Phase transitions in two-dimensional systems with continuous symmetry have attracted considerable attention since their theoretical prediction by Berezinskii, Kosterlitz, and Thouless, providing a new topological ordering through the binding of vortex-antivortex pairs \cite{Berezinskii,Kosterlitz}.
Unlike conventional thermodynamic phase transitions, which are prohibited by the Mermin-Wagner theorem in two-dimensional systems \cite{Coleman,Mermin,Hohenberg}, Berezinskii-Kosterlitz-Thouless (BKT) transitions exhibit a critical line below the BKT transition temperature, $T \leq T^{\rm BKT}$, with the continuously variable critical exponents and the nonzero helicity modulus (superfluid density) showing a discontinuous jump at the BKT transition temperature.
BKT transitions have been observed in $^4$He films \cite{Bishop}, two-dimensional superconductors \cite{Gubser,Hebard,Voss,Wolf,Epstein}, Josephson-junction arrays \cite{Resnick,Voss2}, colloidal crystals \cite{Halperin,Young,Zahn}, and ultracold trapped atomic Bose gases \cite{Hadzibabic}.

An important issue regarding BKT transitions is the relationship between their universality and the topological aspects of the vortices.
Consider the example of a two-dimensional system containing vortices classified by nontrivial discrete groups.
In the case of vortices classified by the finite cyclic group $\mathbb{Z}_2$, which appear in Heisenberg anti-ferromagnets on a triangular lattice \cite{Kawamura-so3}, a vortex becomes its anti-vortex.
It remains a significant open problem whether two-dimensional systems containing $\mathbb{Z}_2$ vortices can show transitions for binding of vortex-(anti)vortex pairs and exhibit nontrivial nonzero quantities, such as helicity modulus.
Another example is a two-dimensional system containing fractionally quantized vortices.
In two-dimensional Bose systems with no internal degrees of freedom, the circulations of vortices are quantized by $2 \pi \hbar / M$ with the particle mass $M$, giving a universal jump of the superfluid number density $\Delta \Upsilon$ at the BKT transition temperature $T^{\rm BKT}$ as
\begin{align}
\Delta \Upsilon = \left( \frac{2 M}{\pi \hbar^2} \right) T^{\rm BKT}.
\label{eq:single-component-universal-relation}
\end{align}
It has been predicted that the relation \eqref{eq:single-component-universal-relation} is changed for superfluid $^3$He \cite{Stein,Korshunov}, two-dimensional Bose mixtures with several atom species \cite{Kobayashi}, and spinor Bose systems \cite{Mukerjee,James}, in which vortices have fractional circulations.
A third example is many-particle systems with short-range interactions.
In three-dimensional systems, a first-order transition occurs from fluid to crystal with spontaneous and simultaneous breakings of translational and rotational symmetries at one transition temperature, giving positional and directional orders.
These two symmetry breakings allow two types of topological defects, namely dislocations and disclinations.
The fluid-crystal transition in three-dimensional systems is predicted to be replaced with a two-step transition at different temperatures in two-dimensional systems with bindings of dislocation-antidislocation and disclination-antidisclination pairs \cite{Halperin,Young,Zahn}.
However, other theoretical and experimental studies have shown the first-order transition as in the case of three-dimensional systems \cite{Naumovets}, and the criteria determining the two-step transitions or the first-order transition is still an open question.

In this paper, we suggest an ultracold atomic spinor Bose gas \cite{Kawaguchi,Yukalov} as a good candidate for investigating the relationship between the properties of the BKT transition and the symmetry of the system.
In many experiments on spinor Bose gases, the linear Zeeman effect can be ignored under the conservation of the total spin in isolated situations, and the quadratic Zeeman effect is very important.
Because the sign of the quadratic Zeeman effect can be manipulated experimentally \cite{Leslie}, both positive and negative quadratic Zeeman effects can be analyzed.
Another important parameter is the spin-dependent coupling constant, which is determined by the atomic species.
For the spin-1 case, there is only one independent coupling constant.
Depending on the sign of the quadratic Zeeman effect and the spin-dependent coupling constant, there exist eight distinct ground states having different manifolds and vortices.
In a two-dimensional system, we can expect a rich variety of BKT transitions depending on the manifold of the ground state.
The other important aspect of a spinor Bose gas for investigating BKT transitions is that we can study two types of superfluid densities: the mass and spin superfluid densities, which are defined by the invariance of the Hamiltonian under a global phase shift, and a global spin rotation.
These superfluid densities give various pieces of information, such as the existence of a BKT transition and details of the universal relation \eqref{eq:single-component-universal-relation}.
These characteristics can be experimentally tested for an ultracold spinor Bose gas in an optical trap sliced by a one-dimensional optical lattice.
In this paper, we numerically calculate the superfluid densities for a spin-1 spinor Bose gas and show that (i) the BKT transition occurs only when the vortices can be classified by the integer group $\mathbb{Z}$ and is absent for $\mathbb{Z}_2$ vortices, (ii) two BKT transitions for the mass and spin superfluid densities occur at different temperatures for a positive quadratic Zeeman effect and at the same temperature for a negative quadratic Zeeman effect, and (iii) the universal relation \eqref{eq:single-component-universal-relation} changes only when multiple kinds of vortices contribute to the BKT transition.

Another important feature of the spinor Bose gas is the spin-singlet pairing of two Bose particles at low temperatures in non-magnetic and partial magnetic states \cite{Kawaguchi,Yukalov,Koashi}.
%In three-dimensional system, it has been predicted that the scalar Bose-Einstein condensation of spin-singlet pairs is possible and more energetically favorable than the spinor condensation.
We also numerically calculate the correlation function of the spin-singlet pair amplitude and find that the quasi off-diagonal long-range order with the algebraic decay of the correlation function emerges at the different temperature lower than the BKT transition temperature and gives a new universal relation of the superfluid density.

This paper is organized as follows.
In Sect. \ref{sec:model}, the model Hamiltonian of our work is shown.
We discuss possible ground states at zero temperature, their manifolds, and vortices, and show our numerical results in Sect. \ref{sec:numerical-result}.
Section \ref{sec:summary} is devoted to a summary and discussion.

\section{Model}
\label{sec:model}

We consider a two-dimensional spin-1 spinor Bose gas in the presence of a quadratic Zeeman field.
The Hamiltonian $\mathcal{H} = \mathcal{H}_{\rm kin} + \mathcal{H}_q + \mathcal{H}_{\rm int}$ consists of a kinetic term $\mathcal{H}_{\rm kin}$, a quadratic Zeeman term $\mathcal{H}_q$, and an interaction term $\mathcal{H}_{\rm int}$, given as
\begin{align}
\begin{split}
& \mathcal{H}_{\rm kin} = \int d^2x\: \sum_{m = -1}^1 \left( \frac{\hbar^2}{2 M} |\nabla \psi_m|^2 \right), \\
& \mathcal{H}_q = \int d^2x\: q \sum_{m = -1}^1 \left( m^2 |\psi_m|^2 \right), \\
& \mathcal{H}_{\rm int} = \frac{1}{2} \int d^2x\: \left( g_0 \rho^2 + g_1 \Vec{S}^2 \right).
\end{split}
\label{eq:Hamiltonian}
\end{align}
We are interested in the thermal BKT transition at finite temperatures, so we use a classical-field approximation which ignores quantum fluctuations.
The classical field for spin-1 Bosons in the magnetic sublevel $m = 0$, $\pm 1$ with particle mass $M$ is denoted as $\psi_m$.
The particle-number density and the spin density are described by $\rho = \psi^\dagger \psi$ and $\Vec{S} = \psi^\dagger \hat{\Vec{s}} \psi$ with the spinor form of the classical field $\psi = ( \psi_1, \psi_0, \psi_{-1} )^T$ and the spin matrix $\hat{\Vec{s}}$ given by $\hat{s}_x = (s_+ + s_+^T) / 2$, $\hat{s}_y = (s_+ - s_+^T) / (2 i)$, and $\hat{s}_z = \mathrm{diag}( 1, 0, -1 )$, where the raising operator $\hat{s}_+$ is given by
\begin{align}
\hat{s}_+ = \begin{pmatrix} 0 & \sqrt{2} & 0 \\ 0 & 0 & \sqrt{2} \\ 0 & 0 & 0 \end{pmatrix}.
\end{align}
The strengths of the quadratic Zeeman effect, spin-independent inter-particle interaction, and spin-dependent inter-particle interaction are given by $q = (g_{\rm L} \mu_{\rm B} B)^2 / E_{\rm hf}$, $g_0 = 4 \pi \hbar^2 (a_0 + 2 a_2) / (3 M)$, and $g_1 = 4 \pi \hbar^2 (a_2 - a_0) / (3 M)$, respectively, where $g_{\rm L}$ is the Land\'e $g$-factor, $\mu_{\rm B}$ is the Bohr magneton, $B$ is the external magnetic field, $E_{\rm hf}$ is the hyperfine energy splitting, and $a_{S = 0, 2}$ is the $s$-wave scattering length in the total spin $S$ channel.
The thermal average $\langle f \rangle$ of the physical observables $f(\psi, \psi^\dagger)$ is defined as
\begin{align}
\langle f \rangle \equiv
\frac{\displaystyle \int \left( \prod_{m=-1}^1 D\psi_m D\psi_m^\ast\right) \: \left( \int \frac{d^2x}{L^2}\: f \right) e^{- \mathcal{H} / T}}
{\displaystyle \int \left( \prod_{m=-1}^1 D\psi_m D\psi_m^\ast\right) \: e^{- \mathcal{H} / T}},
\label{eq:thermal-average}
\end{align}
under the constraint
\begin{align}
\int \frac{d^2x}{L^2}\: \rho = \bar{\rho},
\end{align}
where $T$, $\bar{\rho}$, and $L$ are the temperature, particle number density, and system size, respectively.

In this paper, we mainly discuss two superfluid densities that are finite below the BKT transition temperature.
When both $g_1$ and $q$ are nonzero, the Hamiltonian $\mathcal{H}$ is invariant under the global phase shift $\psi \to e^{i \Delta} \psi$ and the global spin rotation $\psi \to e^{i \hat{s}_z \Delta} \psi$ along the $z$-axis.
The corresponding mass superfluid density and spin superfluid density are, respectively,
\begin{align}
& \Upsilon_{1} = \frac{2 M}{\hbar^2 L^2} \lim_{\Delta \to 0} \frac{F_{1}(\Delta) - F_0}{\Delta^2},
\label{eq:mass-superfluid-density} \\
& \Upsilon_{z} = \frac{2 M}{\hbar^2 L^2} \lim_{\Delta \to 0} \frac{F_{z}(\Delta) - F_0}{\Delta^2},
\label{eq:spin-superfluid-density}
\end{align}
where $F_1$, $F_{z}$, and $F_0$ are free energies obtained from $- T \log Z$ with the partition function $Z = \langle e^{- \mathcal{H}/T} \rangle$.
We use a periodic boundary condition with a phase twist $\psi(x+L,y) =\exp(i \Delta L) \psi(x,y)$ for $F_1$, spin twist $\psi(x+L,y) = \exp(i \hat{s}_z \Delta L) \psi(x,y)$ for $F_z$, and no twist for $F_0$.
Without the quadratic Zeeman effect with $q = 0$, the Hamiltonian $\mathcal{H}$ is also invariant under two additional spin rotations $\psi \to \exp(i \hat{s}_{x,y} \Delta) \psi$ along the $x$ and $y$-axes, and we can further define two additional superfluid densities $\Upsilon_{x,y}$ and we expect $\Upsilon_x = \Upsilon_y = \Upsilon_z$.

We parametrize the strengths of the quadratic Zeeman effect $q$ and the spin-dependent inter-particle interaction $g_1$ as
\begin{align}
q = \tilde{g} \bar{\rho} \sin\upsilon, \quad
g_1 = \tilde{g} \cos\upsilon.
\label{eq:def-theta}
\end{align}
Depending on $\tilde{g}$ and $\upsilon$ in Eq. \eqref{eq:def-theta}, there are eight different types of ground states for the Hamiltonian $\mathcal{H}$ in Eq. \eqref{eq:Hamiltonian}.
In Table \ref{table:manifold}, we summarize these ground states, their manifolds, and fundamental groups.

\begin{table}[tb]
\centering
\begin{tabular}{cc|cc} \hline
$\tilde{g}$ & $\upsilon$ & $\mathcal{M}$ & $\pi_1(\mathcal{M})$ \\ \hline
\vphantom{$\int_1^2$} $\tilde{g} = 0$ & & $\mathrm{S}^5$ & trivial \\ \hline
\vphantom{$\int_1^2$} & $- 180^\circ < \upsilon < - 90^\circ$ & $\mathrm{O}(2)$ & $\mathbb{Z}$ \\
\vphantom{$\int_1^2$} & $ \upsilon = - 90^\circ$ & $\mathrm{S}^3$ & trivial \\
\vphantom{$\int_1^2$} & $- 90^\circ < \upsilon < 0$ & $(\mathrm{S}^1 \times \mathrm{S}^1) / \mathbb{Z}_2$ & $(\mathbb{Z} \times \mathbb{Z}) / 2$ \\
\vphantom{$\int_1^2$} $\tilde{g} > 0$ & $\upsilon = 0$ & $(\mathrm{S}^1 \times \mathrm{S}^2)/\mathbb{Z}_2$ & $\mathbb{Z}/2$ \\
\vphantom{$\int_1^2$} & $0 < \upsilon \leq \upsilon_{\rm P-BA}$ & $\mathrm{S}^1$ & $\mathbb{Z}$ \\
\vphantom{$\int_1^2$} & $\upsilon_{\rm P-BA} < \upsilon < 180^\circ $ & $\mathrm{S}^1 \times \mathrm{S}^1$ & $\mathbb{Z} \times \mathbb{Z}$ \\
\vphantom{$\int_1^2$} & $\upsilon = 180^\circ $ & $\mathbb{R}\mathrm{P}^3$ & $\mathbb{Z}_2$ \\ \hline
\end{tabular}
\caption{
\label{table:manifold}
Dependence of manifolds $\mathcal{M}$ for ground states and their fundamental groups $\pi_1(\mathcal{M})$ on the strength of the quadratic Zeeman effect $q$ and the spin-dependent inter-particle interaction $g_1$ parametrized with $\tilde{g}$ and $\upsilon$ in Eq. \eqref{eq:def-theta}.
Here, $\upsilon_{\rm P-BA} = \cos^{-1}(-1 / \sqrt{5}) \sim 116.6^\circ$ is defined as the boundary between the polar and broken-axisymmetric states.
}
\end{table}

\section{Numerical Results}
\label{sec:numerical-result}

We numerically investigate the thermodynamic properties for the system described by the Hamiltonian \eqref{eq:Hamiltonian}.
The thermal average defined in Eq. \eqref{eq:thermal-average} can be obtained by the cluster Monte Carlo technique with the Wolff algorithm in a periodic and discretized space.
The quadratic Zeeman term $\mathcal{H}_q$ and the interaction term $\mathcal{H}_{\rm int}$ are treated as the external field and can be incorporated in the Wolff algorithm by adding an additional ghost site which connects with all other sites \cite{Dobias}.
The lattice spacing is $0.5 \ell$, where $\ell$ is the healing length $\ell \equiv \hbar / \sqrt{M g_0 \bar{\rho}}$.
For the spin-dependent interaction strength, we use $\tilde{g} = 0.5 g_0$ except for the spherically symmetric state with $\tilde{g} = 0$ discussed in Sect. \ref{subsec:spherically-symmetric}.
We use $\Delta = 0.02 / \ell$ when calculating superfluid densities.

\subsection{Spherically symmetric state for $\tilde{g} = 0$}
\label{subsec:spherically-symmetric}

We start with the simplest case of $\tilde{g} = 0$ in which there is neither a spin-dependent interaction nor a quadratic Zeeman effect.
In this case, all states satisfying $\rho = \bar{\rho}$ can be the ground state, and the system is equivalent to the standard $\mathrm{O}(6)$ model.
The ground-state manifold $\mathcal{M}$ is homeomorphic to the five-dimensional spherical surface $\mathrm{S}^5$.
The topological charge of vortices which appear at finite temperatures can be classified with the fundamental group of the manifold.
The fundamental group is isomorphic to the trivial as $\pi_1(\mathrm{S}^5) \cong 1$ (trivial) and there is no topologically stable vortex in this state.

Because of the equivalence with the $\mathrm{O}(6)$ model, we expect that no phase transition occurs.
\begin{figure}[htb]
\centering
\begin{minipage}{0.492\linewidth}
\centering
  (a) \\
\includegraphics[height=0.95\linewidth]{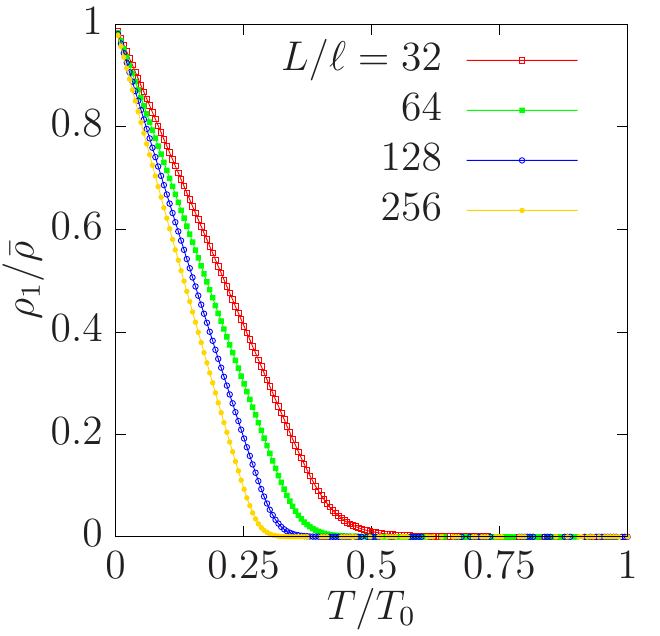}
\end{minipage}
\begin{minipage}{0.492\linewidth}
\centering
  (b) \\
\includegraphics[height=0.95\linewidth]{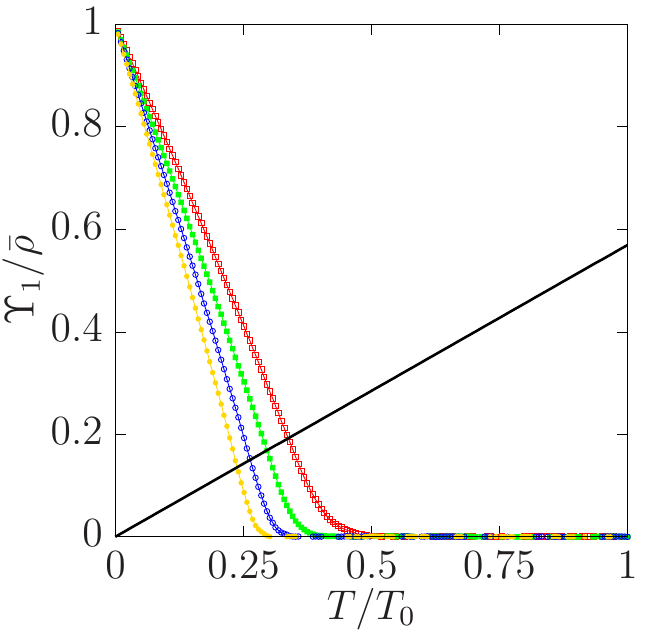}
\end{minipage}
\caption{
\label{fig:O5-rhosm}
Temperature dependence of (a) the mass order parameter $\rho_1$ and (b) the mass superfluid density $\Upsilon_1$ for a spherically symmetric state with $\tilde{g} = 0$.
$T_0$ denotes the BKT transition temperature with the Hamiltonian \eqref{eq:scalar-Hamiltonian} for the scalar Bose system.
The system sizes are $L = 32 \ell$ (red), $L = 64\ell$ (green), $L = 128 \ell$ (blue), and $L = 256 \ell$ (yellow).
The black solid line in panel (b) shows the relation $\Upsilon_1 / T = 2 M / (\pi \hbar^2)$ for the mass superfluid density.
We use the same colors for the system size $L$ in all other figures unless otherwise noted.
}
\end{figure}
To confirm this, we calculate the mass order parameter (condensate density)
\begin{align}
  \rho_1 \equiv G_1(r = L/2),
\end{align}
and the mass superfluid density $\Upsilon_1$.
Here, the mass correlation function
\begin{align}
  G_1(r) \equiv \int \frac{d^2x}{L^2} \int \frac{d \Omega(\Vec{r})}{4 \pi r}\: \langle \psi^\dagger(\Vec{x} + \Vec{r}) \psi(\Vec{x}) \rangle,
  \label{eq:mass-correlation-function}
\end{align}
is obtained by taking an average over the solid angle $\Omega(\Vec{r})$ for the vector $\Vec{r}$.
Figure \ref{fig:O5-rhosm} shows the mass order parameter $\rho_1$ and the mass superfluid density $\Upsilon_1$ as a function of the temperature $T$ normalized by the BKT transition temperature $T_0$ with the Hamiltonian
\begin{align}
\mathcal{H}_{\rm scalar} = 
\int d^2x\: \left( \frac{\hbar^2}{2 M} |\nabla \psi_0|^2 + \frac{g_0}{2} |\psi_0|^4 \right),
\label{eq:scalar-Hamiltonian}
\end{align}
for the scalar Bose system.
Both $\rho_1$ and $\Upsilon_1$ decrease with increasing system size $L$ for the whole temperature regime, and we expect $\rho_1, \Upsilon_1 \to 0$ in the thermodynamic limit $L \to \infty$ at arbitrary temperatures.

\subsection{$\mathrm{SU}(2)$-symmetric state for $\upsilon = -90^\circ$}
\label{subsec:su2-symmetric}

We next consider the case for $\tilde{g} > 0$ and $\upsilon = - 90^\circ$, i.e., where there is only a negative quadratic Zeeman effect.
The ground state satisfies $|\psi_1|^2 + |\psi_{-1}|^2 = \bar{\rho}$ with $\psi_0 = 0$ and is equivalent to the wave function for the spin-$1/2$ system having $\mathrm{SU}(2)$ symmetry.
The ground-state manifold is homeomorphic to the three-dimensional spherical surface $\mathrm{S}^3 \simeq \mathrm{SU}(2)$.
As for the spherically symmetric state, the fundamental group is trivial as $\pi_1(\mathrm{S}^3) \cong 1$ with no topologically stable vortex in this state, and we expect that no phase transition occurs because of its equivalence with the $\mathrm{O}(4)$ model.
\begin{figure}[htb]
\centering
\begin{minipage}{0.492\linewidth}
\centering
  (a) \\
\includegraphics[height=0.95\linewidth]{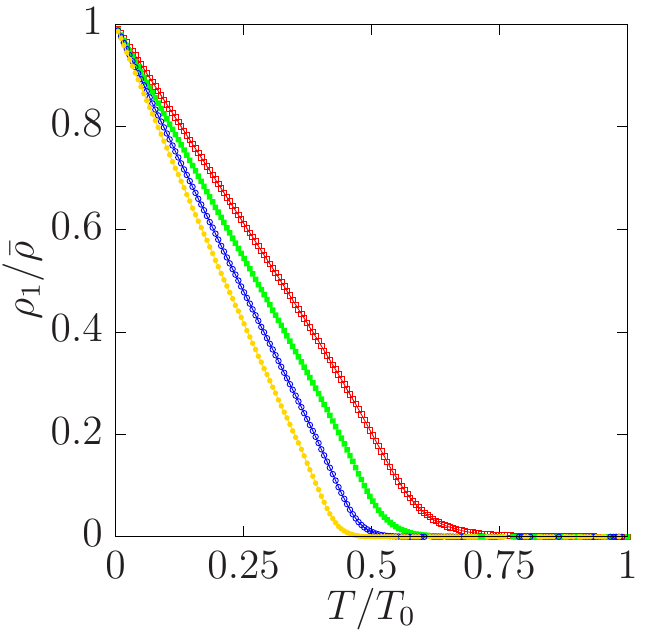}
\end{minipage}
\begin{minipage}{0.492\linewidth}
\centering
  (b) \\
\includegraphics[height=0.95\linewidth]{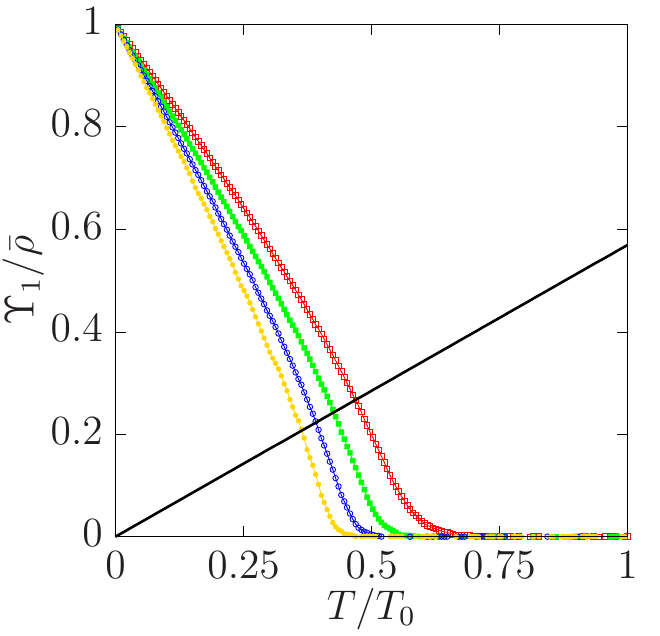}
\end{minipage}
\caption{
\label{fig:su2-rhosm}
Temperature dependence of (a) the mass order parameter $\rho_1$ and (b) the mass superfluid density $\Upsilon_1$ for the $\mathrm{SU}(2)$-symmetric state with $\upsilon = - 90^\circ$.
The black solid line in panel (b) shows the relation $\Upsilon_1 / T = 2 M / (\pi \hbar^2)$ for the mass superfluid density.
}
\end{figure}
Figure \ref{fig:su2-rhosm} shows the dependence of the mass order parameter $\rho_1$ and the mass superfluid density $\Upsilon_1$ on the temperature.
As for the spherically symmetric state, both depend on the system size $L$ for the whole temperature regime, which suggests that there is no phase transition at finite temperatures.

\subsection{Polar state for $0 < \upsilon \leq \upsilon_{\rm P-BA}$}

The nonmagnetic states with $\Vec{S} = 0$ appear as ground states with $\tilde{g} > 0$ and $- 90^\circ < \upsilon \leq \upsilon_{\rm P-BA} \equiv \cos^{-1}(-1/\sqrt{5}) \sim 116.6^\circ$.
The polar state with a positive quadratic Zeeman effect is realized with $\tilde{g} > 0$ and $0 < \upsilon \leq \upsilon_{\rm P-BA}$.
The ground state can be written as 
\begin{align}
\psi = e^{i \varphi} \sqrt{\bar{\rho}} \begin{pmatrix} 0 \\ 1 \\ 0 \end{pmatrix},
\label{eq:polar-+}
\end{align}
The only global phase $\varphi$ contributes to the degree of freedom, and the manifold of the ground state is isomorphic to the one-dimensional circle $\mathrm{S}^1$.
The fundamental group is isomorphic to $\pi_1(\mathrm{S}^1) \cong \mathbb{Z}$.
Because the system has the same symmetry as that of the scalar Bose system, we expect that the essential properties of this state are the same as those of the scalar Bose system.
A typical integer vortex state is expressed with $\varphi = \theta$, where $\theta$ is the angle for the path encircling the vortex, as
\begin{align}
\sqrt{\bar{\rho}} \begin{pmatrix} 0 \\ e^{i \theta} \\ 0 \end{pmatrix}.
\label{eq:polar-+-vortex}
\end{align}

\begin{figure}[htb]
\centering
\begin{minipage}{0.492\linewidth}
\centering
  (a) \\
\includegraphics[height=0.95\linewidth]{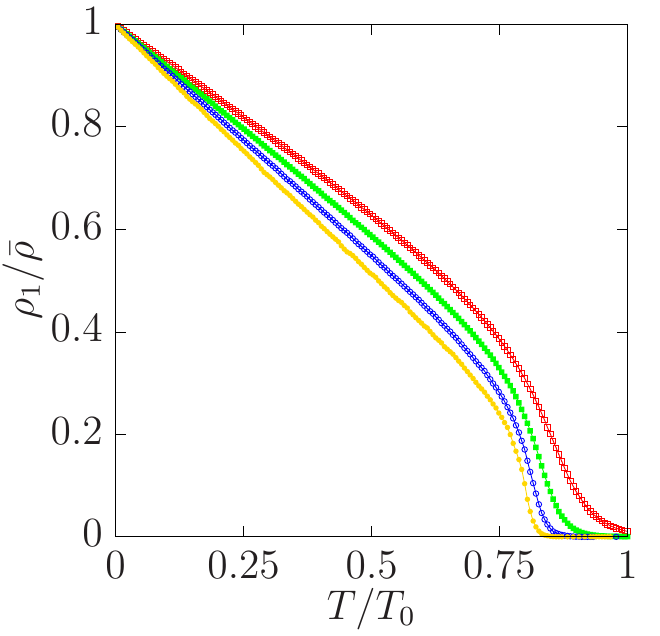}
\end{minipage}
\begin{minipage}{0.492\linewidth}
\centering
  (b) \\
\includegraphics[height=0.95\linewidth]{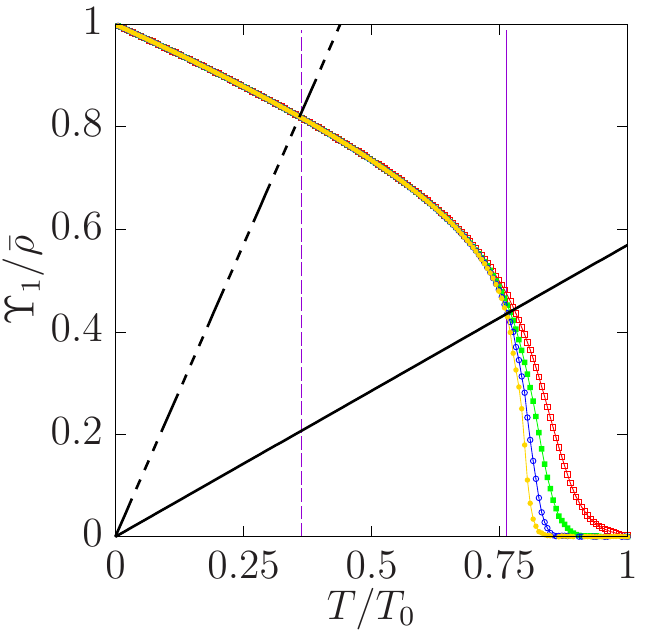}
\end{minipage}
\caption{
\label{fig:u1-rhosm}
Temperature dependence of (a) the mass order parameter $\rho_1$ and (b) the mass superfluid density $\Upsilon_1$ for the polar state with $\upsilon = 45^\circ$.
The black and violet solid lines in panel (b) show the relation $\Upsilon_1 / T = 2 M / (\pi \hbar^2)$ for the mass superfluid density and the mass BKT transition temperature $T^{\rm BKT}_{\rm mass} \approx 0.77 T_0$, respectively.
The black three-dot-chain line and violet dashed line in panel (b) show the relation $\Upsilon_1 / T = 8 M / (\pi \hbar^2)$ for the mass superfluid density and the spin-singlet crossover temperature $T_{\rm singlet}^{\rm CO} \approx 0.36 T_0$, respectively (see Sect. \ref{subsec:singlet-pair}).
}
\end{figure}
Figure \ref{fig:u1-rhosm} shows the dependence of the mass order parameter $\rho_1$ and the mass superfluid density $\Upsilon_1$ with $\upsilon = 45^\circ$.
While the order parameter $\rho_1$ depends on the system size $L$ for the whole temperature regime, as for the two former cases discussed in Sects. \ref{subsec:spherically-symmetric} and \ref{subsec:su2-symmetric}, the mass superfluid density $\Upsilon_1$ has no system size dependence at low temperatures, which suggests a BKT phase with the finite superfluid density $\Upsilon_1$ with vanishing order parameter $\rho_1$.

\begin{figure}[htb]
\centering
\begin{minipage}{0.492\linewidth}
\centering
 (a) \\
\includegraphics[height=0.95\linewidth]{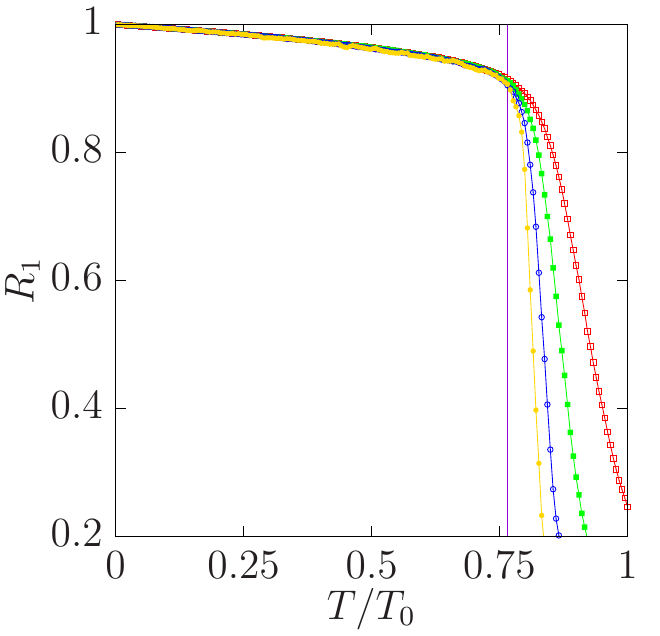}
\end{minipage}
\begin{minipage}{0.492\linewidth}
\centering
 (b) \\
\includegraphics[height=0.95\linewidth]{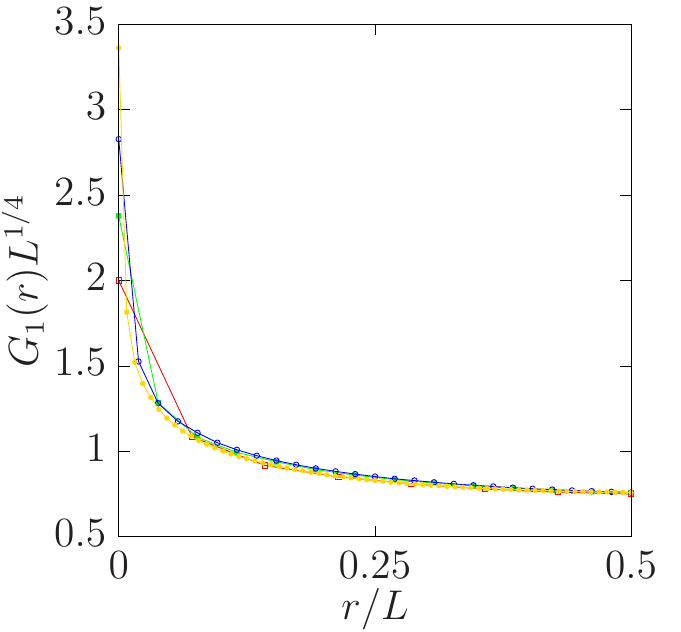}
\end{minipage}
\caption{
\label{fig:u1-correlation}
(a) Temperature dependence of the mass correlation ratio $R_1$ and (b) finite-size scaling of the mass correlation function $G_1$ at the mass BKT transition temperature $T^{\rm BKT}_{\rm mass}$ with the critical exponent $1/4$ for the polar state with $\upsilon = 45^\circ$.
The violet solid line in panel (a) shows the mass BKT transition temperature $T^{\rm BKT}_{\rm mass} \approx 0.77 T_0$.
}
\end{figure}
The mass BKT transition temperature $T^{\rm BKT}_{\rm mass}$, below which the mass superfluid density $\Upsilon_1$ takes a finite value in the thermodynamic limit, can be estimated using the mass correlation function $G_1(r)$ defined in Eq. \eqref{eq:mass-correlation-function}.
The mass correlation ratio $R_1 = G_1(L/4) / G_1(L/2)$ has no system size dependence at $T < T^{\rm BKT}_{\rm mass}$.
Figure \ref{fig:u1-correlation} (a) shows the mass correlation ratio $R_1$, which takes almost the same value at low temperatures and depends on the system size $L$ at $T \geq T^{\rm BKT}_{\rm mass}$.
We here fix the mass BKT transition temperature $T^{\rm BKT}_{\rm mass} \approx 0.77 T_0$ as the temperature at which the standard deviation of four mass correlation ratios $R_1(L = 32\ell, 64\ell, 128\ell, 256\ell)$ becomes 1\%:
\begin{align}
\begin{split}
  & \frac{1}{4 \bar{R}_1} \sqrt{\sum_{i = 1}^4 \{ R_1(L_i) - \bar{R}_1\}^2} = 0.01, \\
  & \bar{R}_1 \equiv \frac{1}{4} \sum_{i=1}^4 R_1(L_i),
\end{split}
\end{align}
where $L_{1,2,3,4} \equiv 32\ell, 64\ell, 128\ell, 256\ell$.
The mass BKT transition temperature $T^{\rm BKT}_{\rm mass}$ can be estimated by the finite-size scaling of the mass correlation function $G_1$.
For a simple BKT transition in the scalar Bose system and $XY$-model, the correlation function $G_1$ obeys the power-law $G_1 \propto r^{- 1/4}$ with the critical exponent $1/4$ at the mass BKT transition temperature $T^{\rm BKT}_{\rm mass}$ in the thermodynamic limit, and we expect the universal form $G_1(r/L) L^{1/4}$ as a function of $r / L$ when the system size $L$ is finite.
Figure \ref{fig:u1-correlation} (b) shows the dependence of $G_1 L^{1/4}$ on $r / L$ at the estimated mass BKT transition temperature $T^{\rm BKT}_{\rm mass}$.
The universality of the mass correlation function $G_1$ is good at large distances $r$, which consolidates the estimation of the mass BKT transition temperature $T^{\rm BKT}_{\rm mass}$ by the mass correlation ratio $R_1$.
At small distances $r$, the correlation function $G_1$ deviates from the universal behavior because of the gapful amplitude mode, which is not negligible at length scales comparable to the healing length $\ell$.

The black solid line in Fig. \ref{fig:u1-rhosm} (b) [and Figs. \ref{fig:O5-rhosm} (b) and \ref{fig:su2-rhosm} (b)] shows the relation $\Upsilon_1 / T = 2 M / (\pi \hbar^2)$.
This line intersects the mass superfluid density $\Upsilon_1$ at almost the BKT transition temperature $T^{\rm BKT}_{\rm mass}$, suggesting that the standard universal relation in Eq. \eqref{eq:single-component-universal-relation} for the mass superfluid density applies:
\begin{align}
  \Delta \Upsilon_1 = \left(\frac{2M}{\pi \hbar^2}\right) T^{\rm BKT}_{\rm mass}.
  \label{eq:mass-universal-relation}
\end{align}

We also examine the spin superfluid density $\Upsilon_{z}$.
\begin{figure}[htb]
\centering
\includegraphics[height=0.5\linewidth]{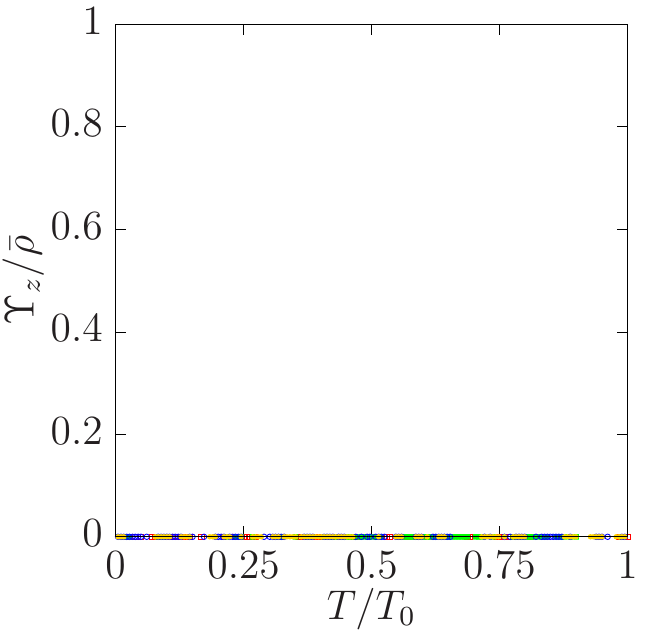}
\caption{
\label{fig:u1-rhoss}
Temperature dependence of the spin superfluid density $\Upsilon_{z}$ for the polar state with $\upsilon = 45^\circ$.
}
\end{figure}
As shown in Fig. \ref{fig:u1-rhoss}, the spin superfluid density $\Upsilon_{z}$ vanishes at all temperatures and the system never shows spin superfluidity because there is no spin degree of freedom in the ground state, as shown in Eq. \eqref{eq:polar-+}.
The absence of the BKT transition in the spin part can also be intuitively explained by the fact that only the global phase contributes to vortices in this state, as shown in Eq. \eqref{eq:polar-+-vortex}. 

\subsection{Anti-ferromagnetic state for $-90^\circ < \upsilon < 0$}

The anti-ferromagnetic state with a negative quadratic Zeeman effect is also a nonmagnetic state with $\Vec{S} = 0$ and is realized for $\tilde{g} > 0$ and $- 90^\circ < \upsilon < 0$.
The ground state can be written as
\begin{align}
\psi = e^{i \varphi} \sqrt{\frac{\bar{\rho}}{2}} \begin{pmatrix} e^{- i \gamma} \\ 0 \\ e^{i \gamma} \end{pmatrix},
\label{eq:polar--}
\end{align}
The manifold of the ground state is isomorphic to $(\mathrm{S}^1 \times \mathrm{S}^1) / \mathbb{Z}_2$, where the discrete $\mathbb{Z}_2$ symmetry corresponds to the equivalence between $(\varphi, \gamma)$ and $(\varphi + \pi, \gamma + \pi)$ in Eq. \eqref{eq:polar--}.
The fundamental group $\pi_1((\mathrm{S}^1 \times \mathrm{S}^1) / \mathbb{Z}_2) \cong (\mathbb{Z} \times \mathbb{Z})/ 2$ can be separately considered as a global phase part $\pi_1(\mathrm{S}^1 / \mathbb{Z}_2) \cong \mathbb{Z} / 2$ and spin part $\pi_1(\mathrm{S}^1 / \mathbb{Z}_2) \cong \mathbb{Z} / 2$.
Two typical half-quantized vortex states are expressed by $\varphi = \gamma = \theta / 2$ as
\begin{align}
\sqrt{\frac{\bar{\rho}}{2}} \begin{pmatrix} 1 \\ 0 \\ e^{i \theta} \end{pmatrix},
\label{eq:polar-half-vortex-1}
\end{align}
and with $\varphi = - \gamma = \theta / 2$ as
\begin{align}
\sqrt{\frac{\bar{\rho}}{2}} \begin{pmatrix} e^{i \theta} \\ 0 \\ 1 \end{pmatrix}.
\label{eq:polar-half-vortex-2}
\end{align}

\begin{figure}[htb]
\centering
\begin{minipage}{0.492\linewidth}
\centering
  (a) \\
\includegraphics[height=0.95\linewidth]{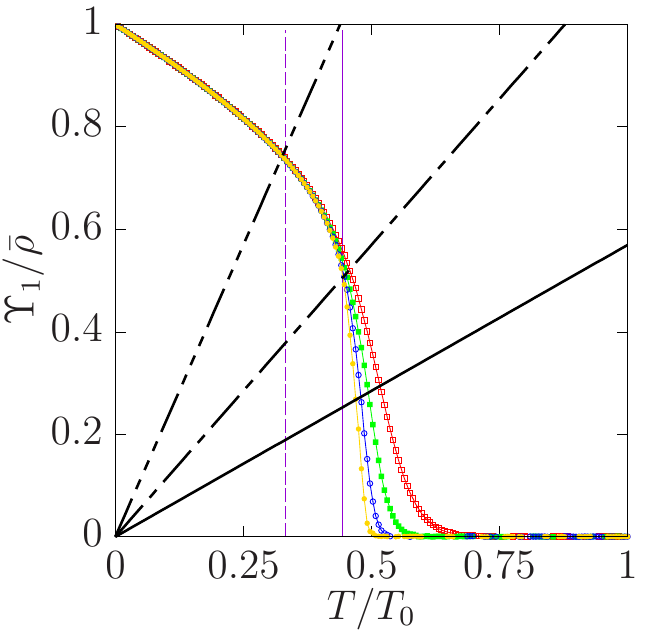}
\end{minipage}
\begin{minipage}{0.492\linewidth}
\centering
  (b) \\
\includegraphics[height=0.95\linewidth]{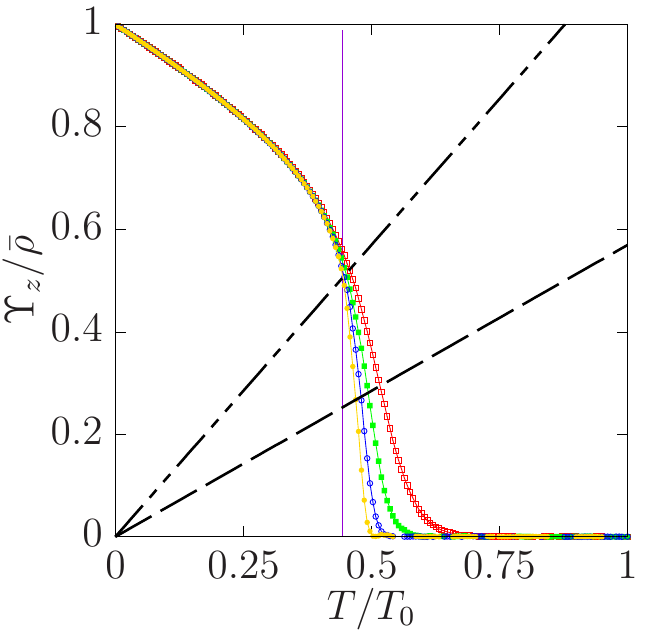}
\end{minipage}
\caption{
\label{fig:u1u12-rhos}
Temperature dependence of (a) the mass superfluid density $\Upsilon_1$ and (b) the spin superfluid density $\Upsilon_z$ for the anti-ferromagnetic state with $\upsilon = -45^\circ$.
The black solid and dash-dot lines in panel (a) show the relations $\Upsilon_1 / T = 2 M / (\pi \hbar^2)$ and $\Upsilon_1 / T = 4 M / (\pi \hbar^2)$, respectively, and the dashed and two-dot chain lines in (b) show $\Upsilon_z / T = 2 M / (\pi \hbar^2)$ and $\Upsilon_z / T = 4 M / (\pi \hbar^2)$.
The violet solid lines show the mass BKT transition temperature $T^{\rm BKT}_{\rm mass} \approx 0.45 T_0$.
The black three-dot-chain line and violet dashed line in panel (a) show the relation $\Upsilon_1 / T = 8 M / (\pi \hbar^2)$ for the mass superfluid density and the spin-singlet crossover temperature $T_{\rm singlet}^{\rm CO} \approx 0.33 T_0$, respectively (see Sect. \ref{subsec:singlet-pair}).
}
\end{figure}
We now skip to show the mass order parameter $\rho_1$ for this state because the essential properties are the same as those for the previous three states, i.e., it depends on the system size $L$ for the whole temperature regime and is expected to vanish in the thermodynamic limit.
Figure \ref{fig:u1u12-rhos} (a) shows the dependence of the mass superfluid density $\Upsilon_1$ on the temperature for the anti-ferromagnetic state with $\upsilon = -45^\circ$.
As for the polar state, the mass BKT transition temperature $T^{\rm BKT}_{\rm mass} \approx 0.45 T_0$ can be estimated by the mass correlation ratio $R_1$ and the finite-size scaling analysis of the mass correlation function $G_1$ with the critical exponent $1/4$.
In Fig. \ref{fig:u1u12-rhos} (a), the mass superfluid density $\Upsilon_1$ does not intersect the black solid line for $\Upsilon_1 / T = 2 M / (\pi \hbar^2)$, but intersects the black dash-dot line for $\Upsilon_1 / T = 4 M / (\pi \hbar^2)$ at the estimated mass BKT transition temperature $T^{\rm BKT}_{\rm mass}$, which suggest a two-times larger universal relation:
\begin{align}
& \Delta \Upsilon_1 = \left( \frac{4 M}{\pi \hbar^2} \right) T^{\rm BKT}_{\rm mass}
\label{eq:large-universal-relation-mass},
\end{align}
rather than the standard universal relation in Eq. \eqref{eq:mass-universal-relation}.

The spin superfluid density $\Upsilon_z$ is almost the same as the mass superfluid density shown in Fig. \ref{fig:u1u12-rhos} (b), suggesting that the spin BKT transition temperature $T^{\rm BKT}_{\rm spin}$ is the same as the mass BKT transition temperature $T^{\rm BKT}_{\rm mass}$, and that the universal relation is given by
\begin{align}
& \Delta \Upsilon_z = \left( \frac{4 M}{\pi \hbar^2} \right) T^{\rm BKT}_{\rm spin}.
\label{eq:large-universal-relation-spin}
\end{align}

The anti-ferromagnetic ground state \eqref{eq:polar--} is equivalent to the ground state of the two-component Bose system with an equal mass and density, where $\varphi - \gamma$ and $\varphi + \gamma$ are the phases of the $\psi_{1}$ and $\psi_{-1}$ components, respectively.
The equal mass and spin superfluid densities can be considered to result from the equal superfluid densities of both components where the quasi off-diagonal long-range orders for phases $\varphi - \gamma$ and $\varphi + \gamma$ for both components can be translated to those for the global and relative phases $\varphi$ and $\gamma$.
Therefore, there is no reason for the difference between the mass and spin superfluid densities $\Upsilon_{1}$ and $\Upsilon_z$ and BKT transition temperatures $T^{\rm BKT}_{\rm mass}$ and $T^{\rm BKT}_{\rm spin}$ because both components have equal mass and density.
%Considering that the superfluid density of each component satisfies the simple universal relation \eqref{eq:single-component-universal-relation}, we can understand the two-times larger universal relations \eqref{eq:large-universal-relation-mass} and \eqref{eq:large-universal-relation-spin} for the mass and spin BKT transitions as resulting from the total contribution of both components.

\subsection{Broken-axisymmetric state for $\upsilon_{\rm P-BA} < \upsilon < 180^\circ$}

The partially magnetized broken-axisymmetric state is realized for $\tilde{g} > 0$ and $\upsilon_{\rm P-BA} < \upsilon < 180^\circ$.
Being different from other states, the ground state explicitly depends on $\upsilon$ as
\begin{align}
\psi = e^{i \varphi} \sqrt{\frac{\bar{\rho}}{8}} \begin{pmatrix} e^{- i \gamma} \sqrt{w_+} \\ \sqrt{2 w_-} \\ e^{i \gamma} \sqrt{w_+} \end{pmatrix}, \quad
w_{\pm} \equiv 2 \pm \tan\upsilon,
\label{eq:BA}
\end{align}
with the partial magnetization $\Vec{S}^2 = \bar{\rho}^2 w_+ w_- / 4 = \bar{\rho}^2 (4 - \tan^2\upsilon) / 4$.
Both the global phase $\varphi$ and the angle $\gamma$ for the spin rotation form the manifold of the one-dimensional circle $\mathrm{S}^1$.
The total manifold of the ground state is homeomorphic to the two-dimensional torus $\mathrm{S}^1 \times \mathrm{S}^1$ and the fundamental group is isomorphic to $\pi_1(\mathrm{S}^1 \times \mathrm{S}^1) \cong \mathbb{Z} \times \mathbb{Z}$.
The typical integer phase vortex state and the integer spin vortex state are expressed as $\varphi = \theta$ and $\gamma = 0$, and $\varphi = 0$ and $\gamma = - \theta$ as 
\begin{align}
& e^{i \theta} \sqrt{\frac{\bar{\rho}}{8}} \begin{pmatrix} \sqrt{w_+} \\ \sqrt{2 w_-} \\ \sqrt{w_+} \end{pmatrix},
\label{eq:BA-phase-vortex} \\
& \sqrt{\frac{\bar{\rho}}{8}} \begin{pmatrix} e^{i \theta} \sqrt{w_+} \\ \sqrt{2 w_-} \\ e^{- i \theta} \sqrt{w_+} \end{pmatrix},
\label{eq:BA-spin-vortex}
\end{align}
respectively.

Besides the mass order parameter $\rho_1$, we can consider the spin order parameter $\rho_S$, defined as
\begin{align}
  \rho_S \equiv G_S(r = L/2),
\end{align}
with the spin correlation function
\begin{align}
  G_S(r) \equiv \int \frac{d^2x}{L^2} \int \frac{d \Omega(\Vec{r})}{4 \pi r}\: \langle \Vec{S}(\Vec{x} + \Vec{r}) \cdot \Vec{S}(\Vec{x}) \rangle.
  \label{eq:spin-correlation-function}
\end{align}
\begin{figure}[htb]
\centering
\begin{minipage}{0.492\linewidth}
\centering
(a) \\
\includegraphics[height=0.95\linewidth]{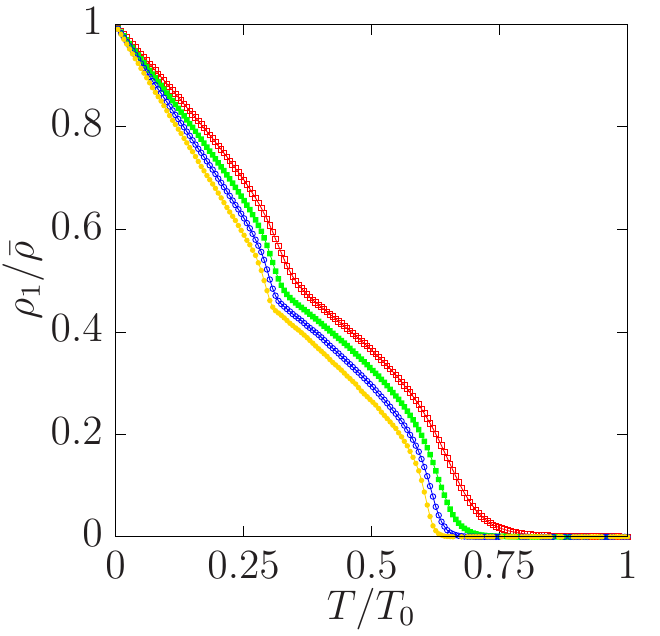}
\end{minipage}
\begin{minipage}{0.492\linewidth}
\centering
(b) \\
\includegraphics[height=0.95\linewidth]{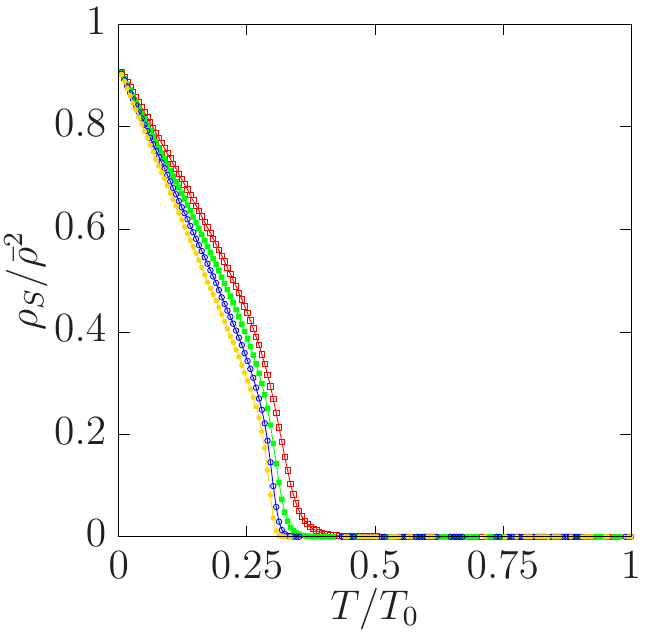}
\end{minipage}
\caption{
\label{fig:BA-rho0}
Temperature dependence of (a) the mass order parameter $\rho_1$ and (b) the spin order parameter $\rho_S$ for the broken-axisymmetric state with $\upsilon = 150^\circ$.
}
\end{figure}
Figure \ref{fig:BA-rho0} shows the mass and spin order parameters $\rho_1$ and $\rho_S$ for the broken-axisymmetric state with $\upsilon = 150^\circ$.
Whereas both behaviors are different at finite system sizes, they are also expected to vanish in the thermodynamic limit.

Because the manifold has two independent $S^1$ spaces in the mass and spin parts, we expect two independent BKT transitions in the two parts.
\begin{figure}[htb]
\centering
\begin{minipage}{0.492\linewidth}
\centering
(a) \\
\includegraphics[height=0.95\linewidth]{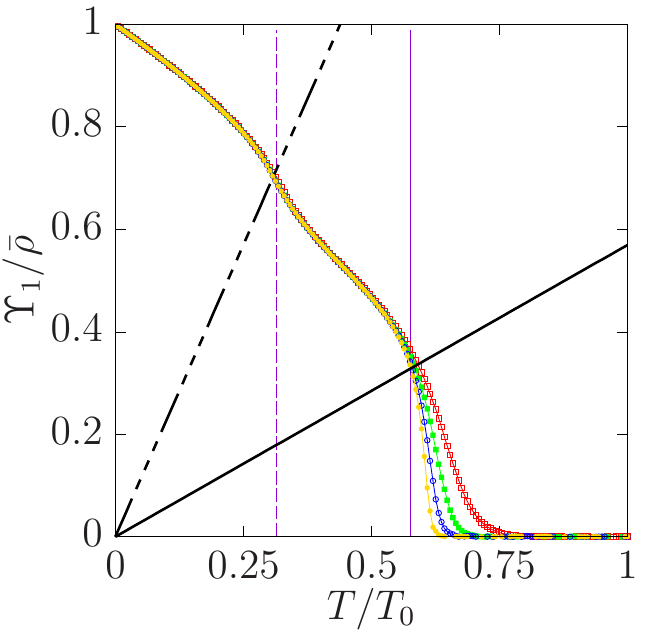}
\end{minipage}
\begin{minipage}{0.492\linewidth}
\centering
(b) \\
\includegraphics[height=0.95\linewidth]{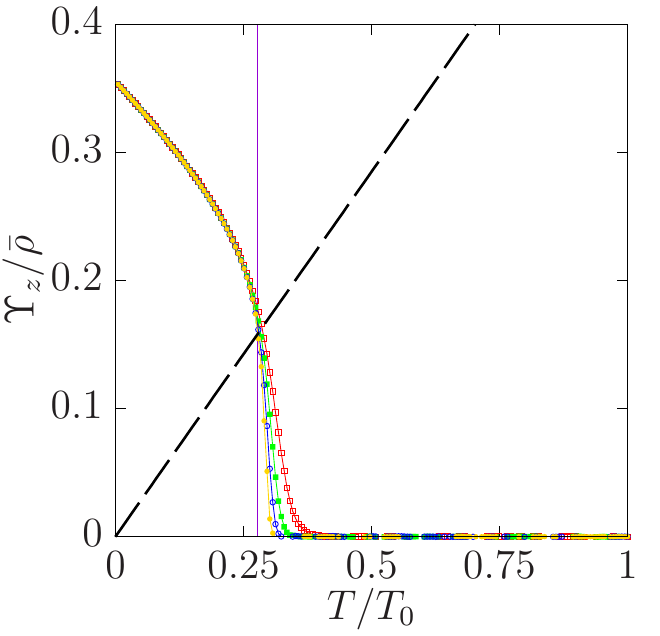}
\end{minipage}
\caption{
\label{fig:BA-rhos}
Temperature dependence of (a) the mass superfluid density $\Upsilon_\rho$ and (b) the spin superfluid density $\Upsilon_{z}$ for the broken-axisymmetric state with $\upsilon = 150^\circ$.
The black solid line in panel (a) shows the relation $\Upsilon_1 / T = 2 M / (\pi \hbar^2)$ and the dashed line in (b) shows $\Upsilon_z / T = 2 M / (\pi \hbar^2)$.
The violet solid line in panel (a) shows the mass BKT transition temperature $T^{\rm BKT}_{\rm mass} \approx 0.58 T_0$ and in (b) it shows the spin BKT transition temperature $T^{\rm BKT}_{\rm spin} \approx 0.28 T_0$.
The black three-dot-chain line and violet dashed line in panel (a) show the relation $\Upsilon_1 / T = 8 M / (\pi \hbar^2)$ for the mass superfluid density and the spin-singlet crossover temperature $T_{\rm singlet}^{\rm CO} \approx 0.32 T_0$, respectively (see Sec. \ref{subsec:singlet-pair}).
}
\end{figure}
Figure \ref{fig:BA-rhos} shows the dependence of the mass and spin superfluid densities $\Upsilon_1$ and $\Upsilon_{z}$ on the temperature for the broken-axisymmetric state with $\upsilon = 150^\circ$.
Both superfluid densities have no system size dependence at low temperatures, suggesting two BKT transitions for the mass and spin parts.
Being different from the anti-ferromagnetic state, the two BKT transition temperatures, $T^{\rm BKT}_{\rm mass}$ for the mass part and $T^{\rm BKT}_{\rm spin}$ for the spin part, are apparently different because the phase $\varphi$ and spin angle $\gamma$ in Eq. \eqref{eq:BA} are completely independent and there is no extra symmetry between them, unlike the extra $\mathbb{Z}_2$ symmetry for the anti-ferromagnetic state.

\begin{figure}[htb]
\centering
\begin{minipage}{0.492\linewidth}
\centering
  (a) \\
\includegraphics[height=0.95\linewidth]{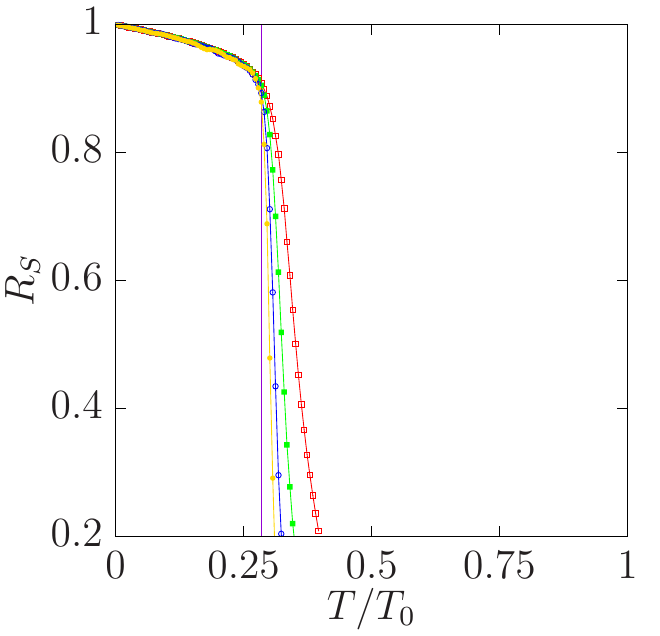}
\end{minipage}
\begin{minipage}{0.492\linewidth}
\centering
 (b) \\
\includegraphics[height=0.95\linewidth]{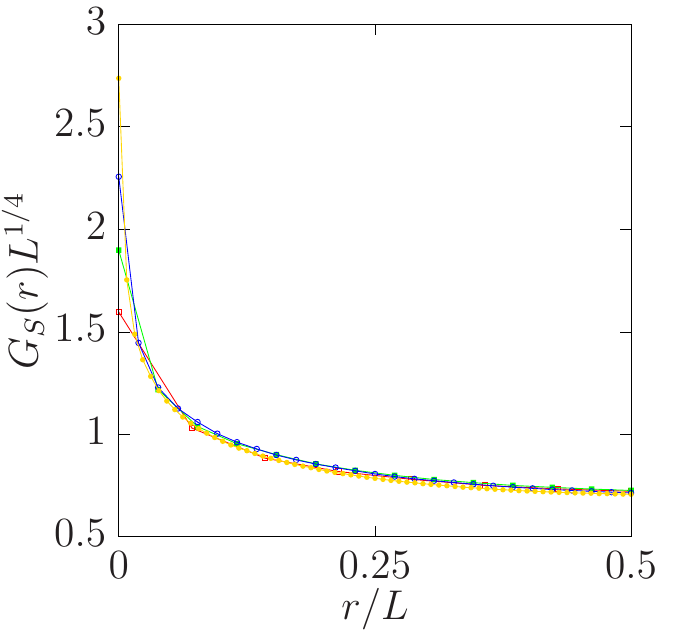}
\end{minipage}
\caption{
\label{fig:BA-correlation}
(a) Temperature dependence of the spin correlation ratio $R_S$ and (b) finite-size scaling of the spin correlation function $G_S$ at the spin BKT transition temperature $T^{\rm BKT}_{\rm spin}$ with the critical exponent $1/4$ for the broken-axisymmetric state with $\upsilon = 150^\circ$.
The violet solid line in panel (a) shows the spin BKT transition temperature $T^{\rm BKT}_{\rm spin} \approx 0.28 T_0$.
}
\end{figure}
The mass BKT transition temperature $T^{\rm BKT}_{\rm mass} \approx 0.58 T_0$ can be estimated from the mass correlation ratio $R_1$ and the finite-size scaling analysis of the mass correlation function $G_1$ with the critical exponent $1/4$.
The spin BKT transition temperature $T^{\rm BKT}_{\rm spin}$ can be estimated by the spin correlation function $G_S(r)$ defined in Eq. \eqref{eq:spin-correlation-function}.
Figure \ref{fig:BA-correlation} (a) shows the spin correlation ratio $R_S = G_S(L/4) / G_S(L/2)$, which depends on the system size $L$ at $T \geq T^{\rm BKT}_{\rm spin} \approx 0.28 T_0$.
The spin BKT transition temperature $T^{\rm BKT}_{\rm spin}$ can be estimated by finite-size scaling of the spin correlation function $G_S$ with the critical exponent $1/4$.
Figure \ref{fig:BA-correlation} (b) shows the dependence of $G_S L^{1/4}$ on $r / L$ at the spin BKT transition temperature $T^{\rm BKT}_{\rm spin}$.
The universality of the spin correlation function $G_S$ is good at large distance $r$.

As shown in Fig \ref{fig:BA-rhos}, both the mass and spin superfluid densities $\Upsilon_1$ and $\Upsilon_z$ satisfy the standard universal relation in Eq. \eqref{eq:mass-universal-relation} and
\begin{align}
  \Delta \Upsilon_z = \left(\frac{2M}{\pi \hbar^2}\right) T^{\rm BKT}_{\rm spin}.
  \label{eq:spin-universal-relation}
\end{align}
We can expect that two independent mass and spin BKT transitions are induced by the phase and spin vortices shown in Eqs. \eqref{eq:BA-phase-vortex} and \eqref{eq:BA-spin-vortex}, respectively.

\subsection{Ferromagnetic state for $-180^\circ < \upsilon < -90^\circ$}

The fully magnetized ferromagnetic state with a negative quadratic Zeeman effect appears as ground states with $\tilde{g} > 0$ and $- 180^\circ < \upsilon < - 90^\circ$.
The ground state is expressed as
\begin{align}
\psi = \frac{\sqrt{\bar{\rho}}}{2} \begin{pmatrix} e^{- i \alpha} (1 + \phi) \\ 0 \\ e^{i \gamma} (1 - \phi) \end{pmatrix},
\label{eq:Ferro--}
\end{align}
where $\phi$ takes $\phi = 1$ or $\phi = -1$, giving a two-fold discrete manifold.
The total manifold of the ground state is constructed by the two-fold discrete group $\mathbb{Z}_2$ for $\phi$ and the one-dimensional circle $\mathrm{S}^1 \simeq \mathrm{SO}(2)$ for spin angles $\alpha$ and $\gamma$, and is homeomorphic to $\mathbb{Z}_2 \ltimes \mathrm{SO}(2) \simeq \mathrm{O}(2)$.
The fundamental group is isomorphic to $\pi_1(\mathrm{O}(2)) \cong \mathbb{Z}$ for both $\phi = 1$ and $\phi = -1$.
The 0th homotopy set also gives the nontrivial group $\pi_0(\mathrm{O}(2)) \cong \mathbb{Z}_2$.
Two typical vortex states are given by inserting $\alpha = \theta$ and $\phi = 1$ as
\begin{align}
\psi = \sqrt{\bar{\rho}} \begin{pmatrix} e^{- i \theta} \\ 0 \\ 0 \end{pmatrix},
\label{eq:Ferro--vortex-1}
\end{align}
and $\gamma = \theta$ and $\beta = \pi$ as
\begin{align}
\psi = \sqrt{\bar{\rho}} \begin{pmatrix} 0 \\ 0 \\ e^{i \theta} \end{pmatrix}.
\label{eq:Ferro--vortex-2}
\end{align}

\begin{figure}[htb]
\centering
\begin{minipage}{0.492\linewidth}
\centering
(a) \\
\includegraphics[height=0.95\linewidth]{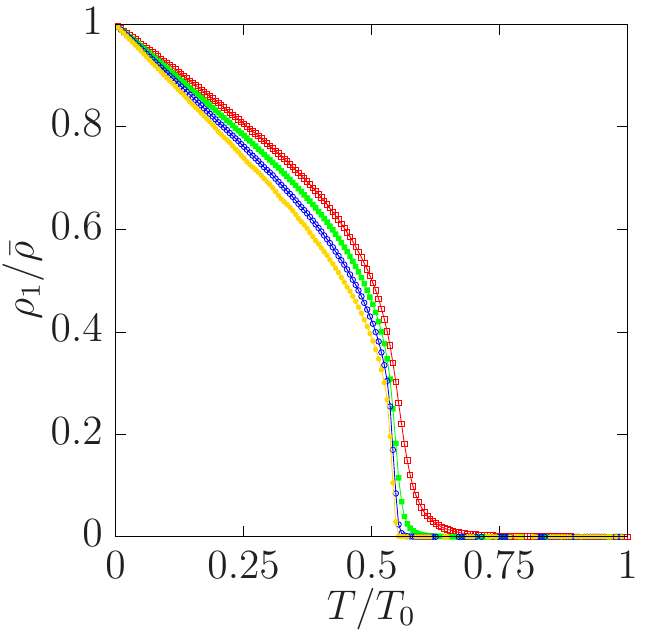}
\end{minipage}
\begin{minipage}{0.492\linewidth}
\centering
(b) \\
\includegraphics[height=0.95\linewidth]{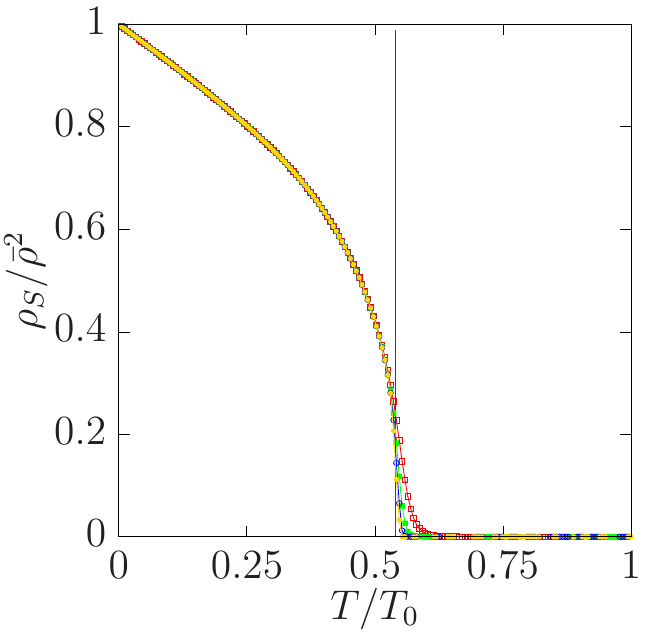}
\end{minipage}
\caption{
\label{fig:ferro--rho0}
Temperature dependence of (a) the mass order parameter $\rho_1$ and (b) the spin order parameter $\rho_S$ for the ferromagnetic state with $\upsilon = -135^\circ$.
The violet solid line in panel (b) shows the spin thermodynamic transition temperature $T^{\rm c}_{\rm spin} \approx 0.54 T_0$.
}
\end{figure}
Because of the two-fold discrete symmetry of the manifold in the spin part, we can expect an Ising-like thermodynamic phase transition.
Figure \ref{fig:ferro--rho0} shows the mass and spin order parameters $\rho_1$ and $\rho_S$ for the ferromagnetic state with $\upsilon = -135^\circ$.
As for all previous cases, the mass order parameter is expected to vanish in the thermodynamic limit.
The spin order parameter, on the other hand, has no system size dependence at low temperatures and takes a finite value in the thermodynamic limit, suggesting a spin thermodynamic transition instead of a spin BKT transition.

\begin{figure}[htb]
\centering
\begin{minipage}{0.492\linewidth}
\centering
(a) \\
\includegraphics[height=0.95\linewidth]{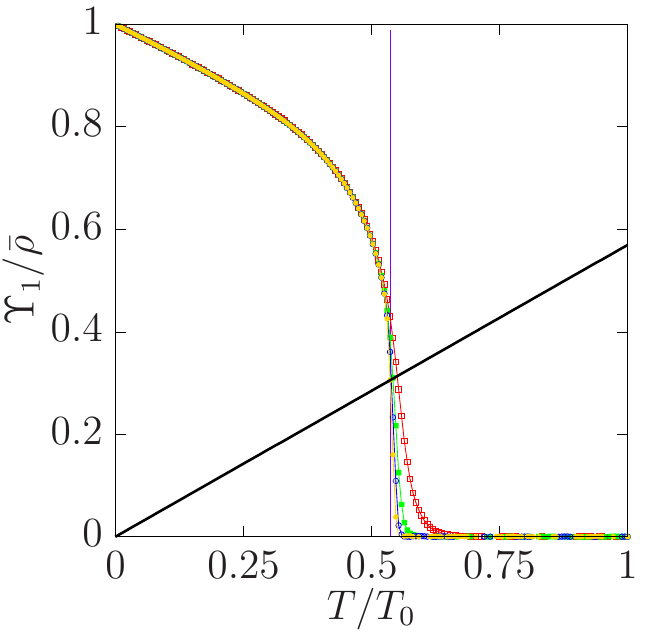}
\end{minipage}
\begin{minipage}{0.492\linewidth}
\centering
(b) \\
\includegraphics[height=0.95\linewidth]{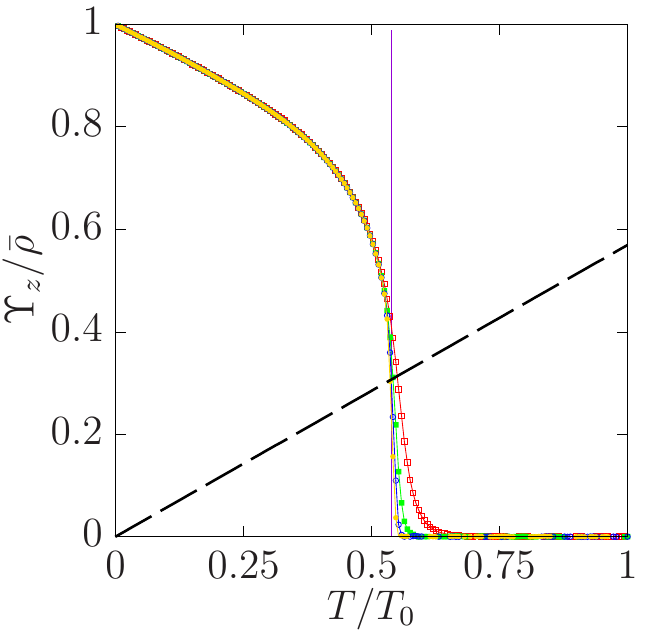}
\end{minipage}
\caption{
\label{fig:ferro--rhos}
Temperature dependence of (a) the mass superfluid density $\Upsilon_1$ and (b) the spin superfluid density $\Upsilon_{z}$ for the ferromagnetic state with $\upsilon = -135^\circ$.
The black solid line in panel (a) shows the relation $\Upsilon_1 / T = 2 M / (\pi \hbar^2)$ and the dashed line in (b) shows $\Upsilon_z / T = 2 M / (\pi \hbar^2)$.
The violet solid line in panel (a) shows the mass BKT transition temperature $T^{\rm BKT}_{\rm mass} \approx 0.54 T_0$ and in (b) it shows the spin thermodynamic transition temperature $T^{\rm c}_{\rm spin} \approx 0.54 T_0$.
}
\end{figure}
Figure \ref{fig:ferro--rhos} shows the dependence of the mass and spin superfluid densities $\Upsilon_1$ and $\Upsilon_{z}$ on the temperature for the ferromagnetic state with $\upsilon = - 135^\circ$.
Both superfluid densities have no system size dependence at low temperatures.
Whereas the mass superfluid density $\Upsilon_1$ is expected to arise from a BKT transition, the finite spin superfluid density $\Upsilon_z$ is expected to be caused by a thermodynamic phase transition with a finite spin order parameter $\rho_S$, as in the case of superfluidity in a three-dimensional system.
The spin thermodynamic transition temperature $T^{\rm c}_{\rm spin} \approx 0.54 T_0$ can be estimated by the spin correlation ratio $R_S$ and is very close to the estimated mass BKT transition temperature $T^{\rm BKT}_{\rm mass} \approx 0.54$.
As shown in Fig \ref{fig:BA-rhos}, both the mass and spin superfluid densities $\Upsilon_1$ and $\Upsilon_z$ satisfy the standard universal relation in Eq. \eqref{eq:mass-universal-relation} and \eqref{eq:spin-universal-relation}, respectively.
We can expect that the mass BKT transitions are induced by the phase vortex \eqref{eq:Ferro--vortex-1} and \eqref{eq:Ferro--vortex-2} in the broken phase of the discrete symmetry.

We note that in many experiments, the total spin is conserved and discrete symmetry breaking does not occur.
Instead of discrete symmetry breaking, spatial phase separation between $\phi=1$ and $\phi=-1$ in Eq. \eqref{eq:Ferro--} can be expected at the transition temperature $T^{\rm c}_{\rm spin}$.

For the present numerical accuracy, we cannot judge whether these two transition temperatures, $T^{\rm BKT}_{\rm mass}$ and $T^{\rm c}_{\rm spin}$, are different or not and whether the universality class of the spin thermodynamic phase transition is the same as the Ising one.
To address these issues, more detailed numerical work is required with a simpler model having the same manifold [see Eq. \eqref{eq:semi-direct-product}].

\subsection{Polar state with $q = 0$ for $\upsilon = 0^\circ$}

For the nonmagnetic polar state without a quadratic Zeeman effect, i. e., $\tilde{g} > 0$ and $\upsilon = 0$, the ground state can be written as
\begin{align}
\psi = e^{i \varphi} \sqrt{\frac{\bar{\rho}}{2}} \begin{pmatrix} - e^{- i \gamma} \sin\beta \\ \sqrt{2} \cos\beta \\ e^{i \gamma} \sin\beta \end{pmatrix},
\label{eq:polar-0}
\end{align}
The global phase $\varphi$ forms the manifold of the one-dimensional circle $\mathrm{S}^1$.
The spin angles $\beta$ and $\gamma$ satisfy $0 \leq \beta < \pi$ and $- \pi < \gamma \leq \pi$, and form the manifold of the two-dimensional spherical surface $\mathrm{S}^2$.
The total manifold of the ground state is homeomorphic to $(\mathrm{S}^1 \times \mathrm{S}^2) / \mathbb{Z}_2$, where the discrete $\mathbb{Z}_2$ symmetry corresponds to the equivalence between $(\varphi, \beta)$ and $(\varphi + \pi, \beta + \pi)$ in Eq. \eqref{eq:polar-0}.
The fundamental group $\pi_1((\mathrm{S}^1 \times \mathrm{S}^2) / \mathbb{Z}_2) \cong \mathbb{Z} / 2$ can be separately considered as a global-phase part $\pi_1(\mathrm{S}^1 / \mathbb{Z}_2) \cong \mathbb{Z} / 2$ and a spin-angle part $\pi_1(\mathrm{S}^2 / \mathbb{Z}_2) \cong \pi_1(\mathbb{R}\mathrm{P}^2) \cong \mathbb{Z}_2$.
A typical half-quantized vortex state is expressed with $\varphi = \theta / 2$ and $\beta = \theta / 2$ as
\begin{align}
\frac{\sqrt{\bar{\rho}}}{2 \sqrt{2}} \begin{pmatrix}
- e^{- i (\gamma - \pi / 2)} \left( e^{i \theta} - 1 \right) \\
\sqrt{2} \left( e^{i \theta} + 1 \right) \\
e^{i (\gamma + \pi / 2)} \left( e^{i \theta} - 1 \right)
\end{pmatrix}.
\label{eq:polar-half-vortex-0}
\end{align}

\begin{figure}[htb]
\centering
\begin{minipage}{0.492\linewidth}
\centering
(a) \\
\includegraphics[height=0.95\linewidth]{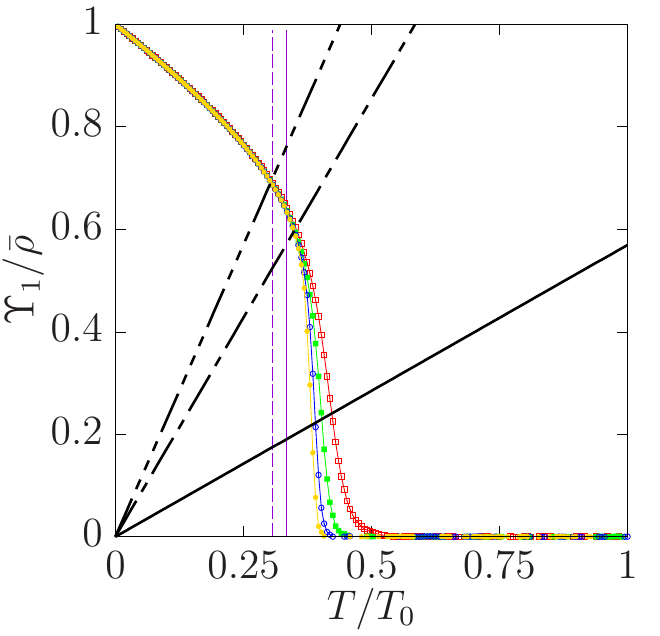}
\end{minipage}
\begin{minipage}{0.492\linewidth}
\centering
(b) \\
\includegraphics[height=0.95\linewidth]{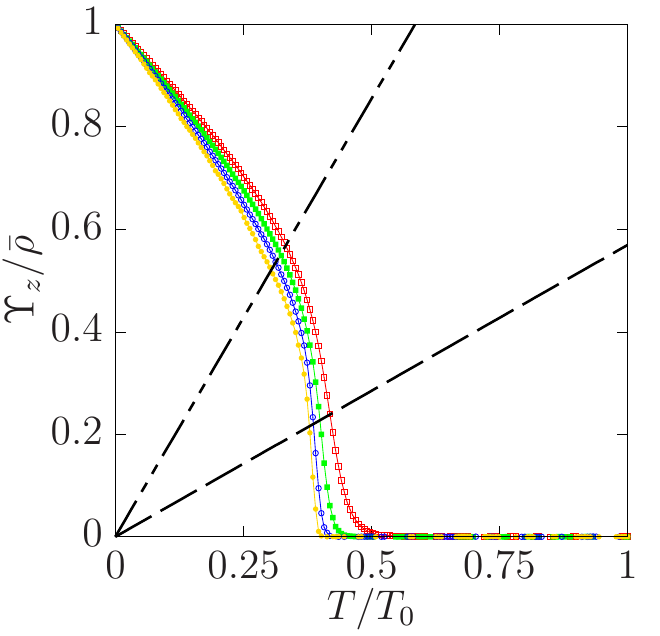}
\end{minipage}
\caption{
\label{fig:polar0-rhos}
Temperature dependence of (a) the mass superfluid density $\Upsilon_1$ and (b) the spin superfluid density $\Upsilon_z$ for the polar state with $\upsilon = 0^\circ$.
The black solid and dash-dot lines in panel (a) show the relations $\Upsilon_1 / T = 2 M / (\pi \hbar^2)$ and $\Upsilon_1 / T = 6 M / (\pi \hbar^2)$, respectively, and the dashed and two-dot chain lines in (b) show $\Upsilon_z / T = 2 M / (\pi \hbar^2)$ and $\Upsilon_z / T = 6 M / (\pi \hbar^2)$.
The violet solid line in panel (a) shows the mass BKT transition temperature $T^{\rm BKT}_{\rm mass} \approx 0.34 T_0$.
The black three-dot-chain line and violet dashed line in panel (a) show the relation $\Upsilon_1 / T = 8 M / (\pi \hbar^2)$ for the mass superfluid density and the spin-singlet crossover temperature $T_{\rm singlet}^{\rm CO} \approx 0.31 T_0$, respectively (see Sec. \ref{subsec:singlet-pair}).
}
\end{figure}
Figure \ref{fig:polar0-rhos} (a) shows the dependence of the mass superfluid density $\Upsilon_1$ on the temperature for the polar state with $\upsilon = 0^\circ$.
As for the polar state with a positive quadratic Zeeman effect and an anti-ferromagnetic state, the mass BKT transition is expected to exhibit a finite mass superfluid density $\Upsilon_1$ with mass BKT transition temperature $T^{\rm BKT}_{\rm mass} \approx 0.34 T_0$, which can be estimated by the mass correlation ratio $R_1$ and the finite-size scaling analysis of the mass correlation function $G_1$ with the critical exponent $1/4$.
The mass superfluid density $\Upsilon_1$ does not support the standard universal relation in Eq. \eqref{eq:mass-universal-relation} but shows a three-times large universal relation
\begin{align}
& \Delta \Upsilon_1 = \left( \frac{6 M}{\pi \hbar^2} \right) T^{\rm BKT}_{\rm mass}
\label{eq:large0-universal-relation-mass},
\end{align}
because the mass superfluid density $\Upsilon_1$ intersects the thick dash-dot line for $\Upsilon_1 / T = 6 M / (\pi \hbar^2)$ at the estimated mass BKT transition temperature $T^{\rm BKT}_{\rm mass}$ as shown in Fig. \ref{fig:polar0-rhos} (a).
As for the anti-ferromagnetic state, the larger universal relation in Eq. \eqref{eq:large0-universal-relation-mass} can be simply understood: all three vortices in Eqs. \eqref{eq:polar-+-vortex}, \eqref{eq:polar-half-vortex-1}, and \eqref{eq:polar-half-vortex-2} contribute to the mass BKT transition.
In this sense, we can understand that the large universal relations in Eqs. \eqref{eq:large-universal-relation-mass} and \eqref{eq:large0-universal-relation-mass} are determined by how many kinds of vortices contribute to the BKT transition rather than the ``fractionalized circulation'' of vortices, i.e., if the fractional circulation determines the universal relation, the two universal relations for the anti-ferromagnetic state and the polar state without the quadratic Zeeman effect would be the same because vortices in these two states have the same mass circulation.

In contrast to the mass superfluid density $\rho_1$, the spin superfluid density $\rho_z$ decreases with increasing system size $L$ for the whole temperature regime, as for $\rho_1$ in the spherically symmetric state in Fig. \ref{fig:O5-rhosm} (b) and the $\mathrm{SU}(2)$-symmetric state in Fig. \ref{fig:su2-rhosm} (b), which suggests the absence of a spin BKT transition.
The topological charge of vortices in the spin space is $\mathbb{Z}_2$, and our result suggests that $\mathbb{Z}_2$ vortices in the manifold of the two-dimensional projective plane $\mathbb{R}\mathrm{P}^2$ do not contribute to the BKT transition.
In other words, the dimension of the low-energy excitations, rather than the existence of vortices, seems to be a more important determinant for the existence of the BKT transition.
The dimension of the low-energy spin excitation is two for $\beta$ and $\gamma$ in Eq. \eqref{eq:polar-0}, the same as that for the classical Heisenberg model that does not show the BKT transition.

\subsection{Ferromagnetic state with $q = 0$ for $\upsilon = 180^\circ$}

For the fully magnetized ferromagnetic state without the quadratic Zeeman effect, i.e., $\tilde{g} > 0$ and $\upsilon = 180^\circ$, the ground state can be written as
\begin{align}
\psi = e^{- i \alpha} \sqrt{\frac{\bar{\rho}}{2}} \begin{pmatrix} e^{- i \gamma} \sqrt{2} \cos^2(\beta/2) \\ \sin\beta \\ e^{i \gamma} \sqrt{2} \sin^2(\beta/2) \end{pmatrix},
\label{eq:Ferro-0}
\end{align}
The spin-rotation angles $\alpha$, $\beta$, and $\gamma$ satisfy $- \pi < \alpha \leq \pi$, $0 \leq \beta < \pi$, and $- \pi < \gamma \leq \pi$, and are equivalent to Euler angles constructed with $\mathrm{SO}(3)$ symmetry.
The manifold of the ground state is homeomorphic to the three-dimensional projective plane $\mathbb{R}\mathrm{P}^3 \simeq \mathrm{SO}(3)$ and the fundamental group is isomorphic to $\pi_1(\mathbb{R}\mathrm{P}^3) \cong \mathbb{Z}_2$.
A typical $\mathbb{Z}_2$ vortex state is expressed with $\alpha = 0$, $\beta = \mathrm{const}$, and $\gamma = - \theta$ as
\begin{align}
\sqrt{\frac{\bar{\rho}}{2}} \begin{pmatrix} e^{i \theta} \sqrt{2} \cos^2(\beta/2) \\ \sin\beta \\ e^{- i \theta} \sqrt{2} \sin^2(\beta/2) \end{pmatrix}.
\label{eq:ferro-vortex}
\end{align}

\begin{figure}[htb]
\centering
\begin{minipage}{0.492\linewidth}
\centering
(a) \\
\includegraphics[height=0.95\linewidth]{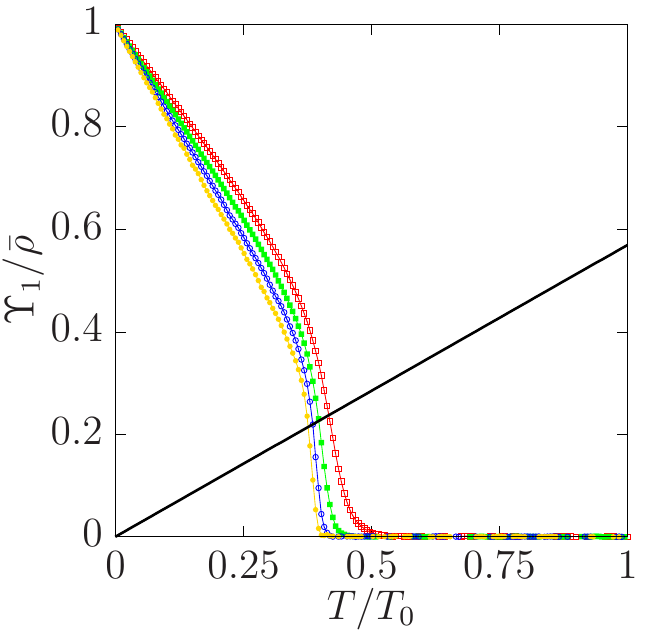}
\end{minipage}
\begin{minipage}{0.492\linewidth}
\centering
(b) \\
\includegraphics[height=0.95\linewidth]{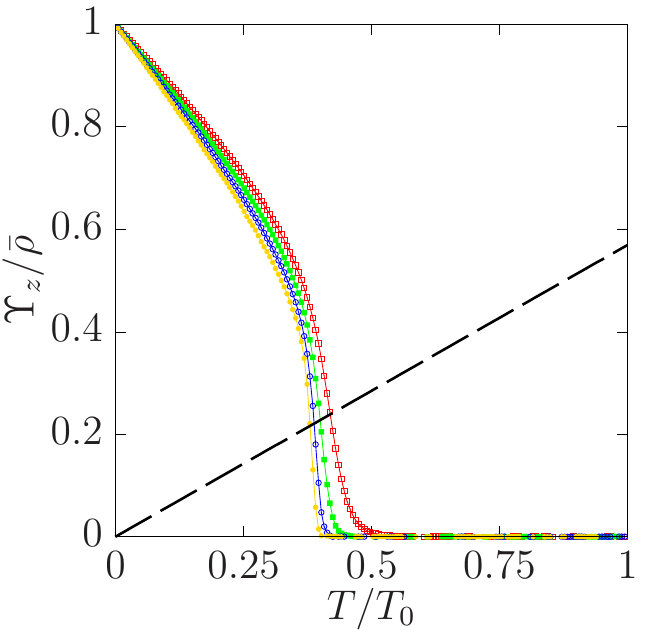}
\end{minipage}
\caption{
\label{fig:Ferro0-rhos}
Temperature dependence of (a) the mass superfluid density $\Upsilon_1$ and (b) the spin superfluid density $\Upsilon_z$ for the ferromagnetic state with $\upsilon = 180^\circ$.
The black solid line in panel (a) shows the relation $\Upsilon_1 / T = 2 M / (\pi \hbar^2)$ and the dashed line in (b) shows $\Upsilon_z / T = 2 M / (\pi \hbar^2)$.
}
\end{figure}
Figure \ref{fig:Ferro0-rhos} shows the dependence of the mass and spin superfluid densities $\Upsilon_1$ and $\Upsilon_{z}$ on the temperature for the ferromagnetic state with $\upsilon = 180^\circ$.
As for the other superfluid densities shown in Figs. \ref{fig:O5-rhosm} (b), \ref{fig:su2-rhosm} (b), and \ref{fig:polar0-rhos} (b), both the mass and spin superfluid densities $\Upsilon_1$ and $\Upsilon_z$ are expected to vanish in the thermodynamic limit, suggesting that there is no transition in this state.
This result suggests that $\mathbb{Z}_2$ vortices do not induce the BKT transition and consolidates the absence of the spin BKT transition for the polar state without the quadratic Zeeman effect where the spin part of the topological charge of vortices is $\mathbb{Z}_2$.

\subsection{Quasi long-range order for spin-singlet pairing}
\label{subsec:singlet-pair}

We here consider the situation in which two Bosons form a spin-singlet pair \cite{Kawaguchi,Yukalov,Koashi},
\begin{align}
\hat{A}_{20}^\dagger = \sum_{m = -1}^1 (-1)^m \hat{\psi}_m^\dagger \hat{\psi}_{-m}^\dagger = \hat{\psi}_0^{\dagger\: 2} - 2 \hat{\psi}_1^\dagger \hat{\psi}_1^\dagger,
\end{align}
  where $\hat{\psi}_m^\dagger$ and $\hat{A}_{20}^\dagger$ are the creation operators for a Boson with the magnetic sublevel $m$ and a spin-singlet pair, respectively.
In a three-dimensional system, it has been predicted that a Bose-Einstein condensed state of spin-singlet pairs:
\begin{align}
(\hat{A}_{20}^\dagger)^N |\mathrm{vacuum}\rangle,
\end{align}
is more energetically favorable than the single-particle condensed state:
\begin{align}
\begin{split}
& \left( - \frac{\hat{\psi}_1}{\sqrt{2}} e^{- i \gamma} \sin\beta + \hat{\psi}_0 \cos\beta + \frac{\hat{\psi}_{-1}}{\sqrt{2}} e^{i \gamma} \sin\beta \right)^N \\
&\quad \times |\mathrm{vacuum}\rangle,
\end{split}
\end{align}
due to the quantum fluctuation at $q = 0$, $g_1 > 0$, and $T = 0$.
However, it has still been an open problem how the paired condensation breaks for finite $q$ and $T$.

In a two-dimensional system, we can expect the quasi off-diagonal long-range order of the spin-singlet pairing, and we here study how it emerges and grows at low temperatures.
In a framework of the classical-field approximation, we can consider the spin-singlet pair amplitude
\begin{align}
A_{20} = \sum_{m=-1}^1 (-1) \psi_m \psi_{-m} = \psi_0^2 - 2 \psi_1 \psi_1,
\end{align}
instead of $\hat{A}_{20}$.
Because the spin-singlet pair amplitude satisfies $|A_{20}|^2 = \rho^2 - \Vec{S}^2$, it becomes finite for the polar, anti-ferromagnetic, and broken-axisymmetric ground states at $T = 0$.
\begin{figure}[htb]
\centering
\begin{minipage}{0.492\linewidth}
\centering
(a) \\
\includegraphics[height=0.95\linewidth]{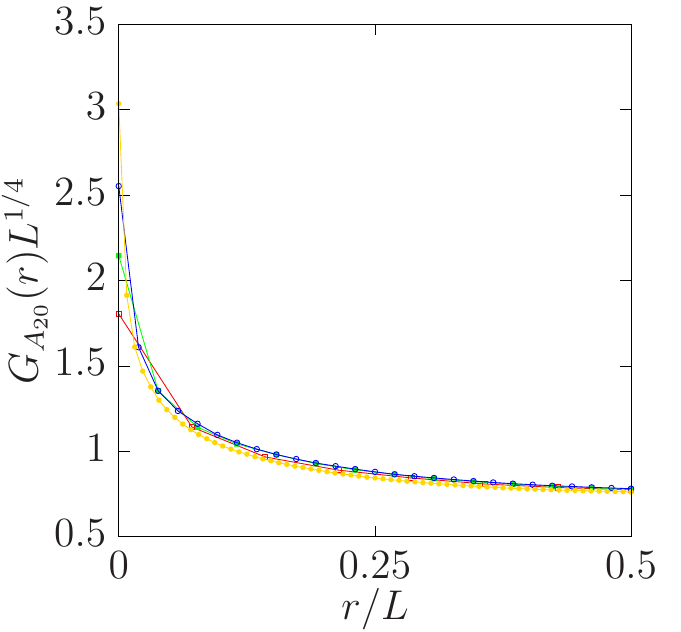}
\end{minipage}
\begin{minipage}{0.492\linewidth}
\centering
(b) \\
\includegraphics[height=0.95\linewidth]{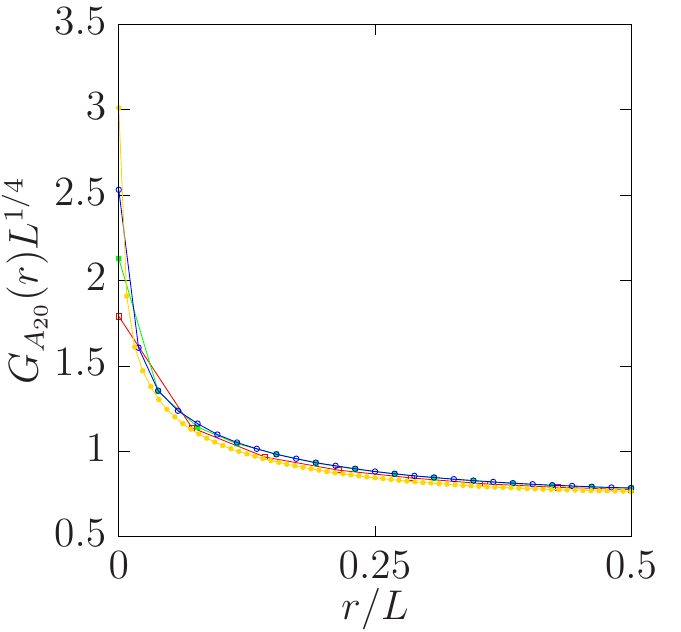}
\end{minipage} \\[10pt]
\begin{minipage}{0.492\linewidth}
\centering
(c) \\
\includegraphics[height=0.95\linewidth]{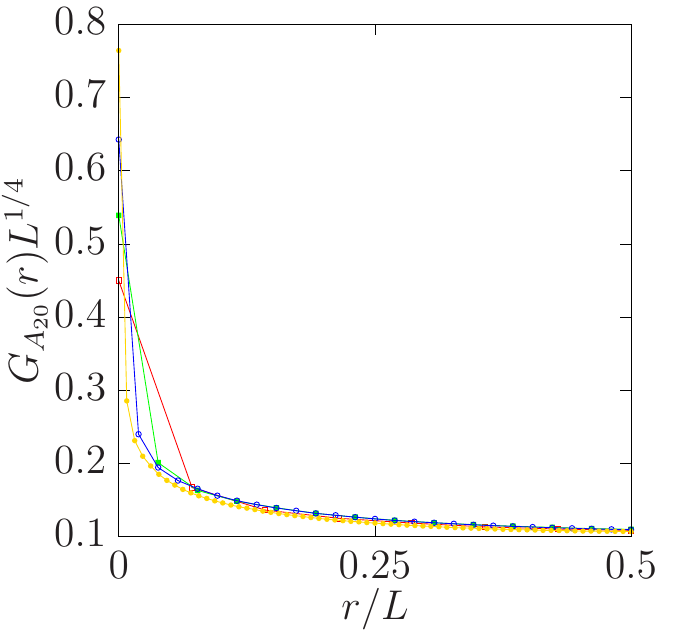}
\end{minipage}
\begin{minipage}{0.492\linewidth}
\centering
(d) \\
\includegraphics[height=0.95\linewidth]{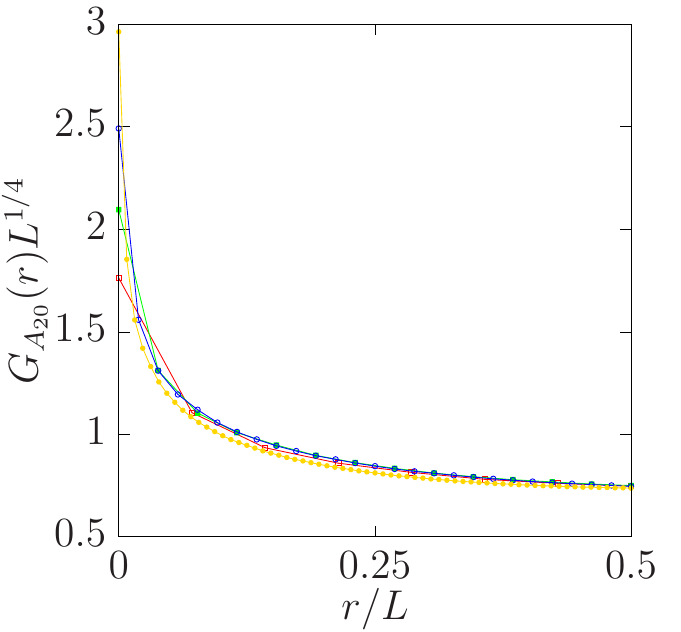}
\end{minipage}
\caption{
\label{fig:correlation-A20}
Finite-size scaling of the spin-singlet pairing correlation function $G_{A_{20}}$ for (a) the polar state with $\upsilon = 0^\circ$, (b) the polar state with $\upsilon = 45^\circ$, (c) the broken-axisymmetric state with $\upsilon = 150^\circ$, (d) and the anti-ferromagnetic state with $\upsilon = -45^\circ$ at the spin-singlet crossover temperature (a) $T_{\rm singlet}^{\rm CO} = 0.34 T_0$, (b) $T_{\rm singlet}^{\rm CO} = 0.36 T_0$, (c) $T_{\rm singlet}^{\rm CO} = 0.32 T_0$, and (d) $T_{\rm singlet}^{\rm CO} = 0.33 T_0$ with the critical exponent $1/4$.}
\end{figure}

Figure \ref{fig:correlation-A20} shows the dependence of $G_{A_{20}} L^{1/4}$ on $r/L$, where the spin-singlet paring correlation function $G_{A_{20}}(r)$ is defined as
\begin{align}
G_{A_{20}}(r) = \int \frac{d^2 x}{L^2} \int \frac{d \Omega(\Vec{r})}{4 \pi r} \langle A_{20}^\ast(\Vec{x} + \Vec{r}) A_{20}(\Vec{x}) \rangle.
\end{align}
The universality of the spin-singlet paring correlation function $G_{A_{20}}$ is good, which suggests the quasi off-diagonal long-range order of the spin-singlet pairing with the algebraic decay of the correlation function $G_{A_{20}} \propto r^{-\eta}$ in the thermodynamic limit.
However, the temperature at which the spin-singlet pairing correlation function satisfies $G_{A_{20}}(r) \propto r^{-1/4}$ is always lower than the mass BKT transition temperature $T_{\rm mass}^{\rm BKT}$ at which the mass correlation function satisfies $G_1(r) \propto r^{-1/4}$.
\begin{figure}[htb]
\centering
\begin{minipage}{0.492\linewidth}
\centering
(a) \\
\includegraphics[height=0.95\linewidth]{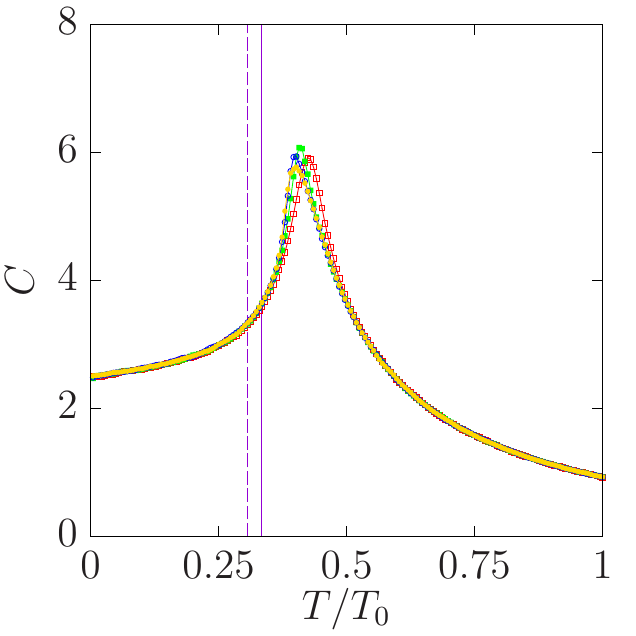}
\end{minipage}
\begin{minipage}{0.492\linewidth}
\centering
(b) \\
\includegraphics[height=0.95\linewidth]{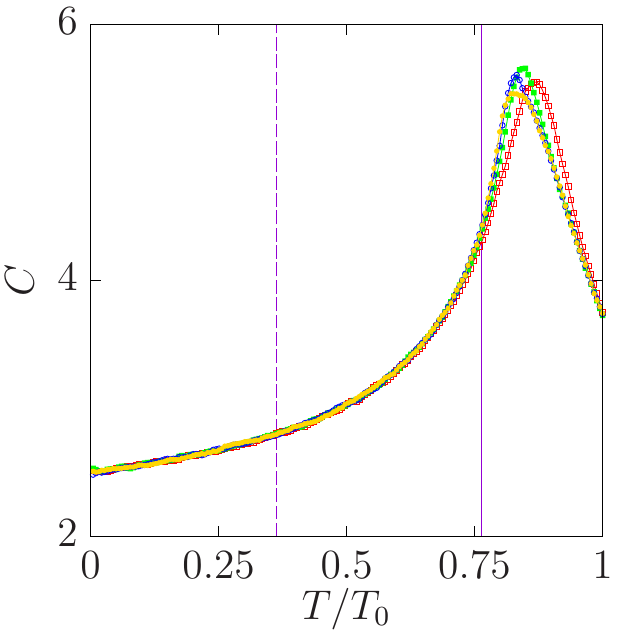}
\end{minipage} \\[10pt]
\begin{minipage}{0.492\linewidth}
\centering
(c) \\
\includegraphics[height=0.95\linewidth]{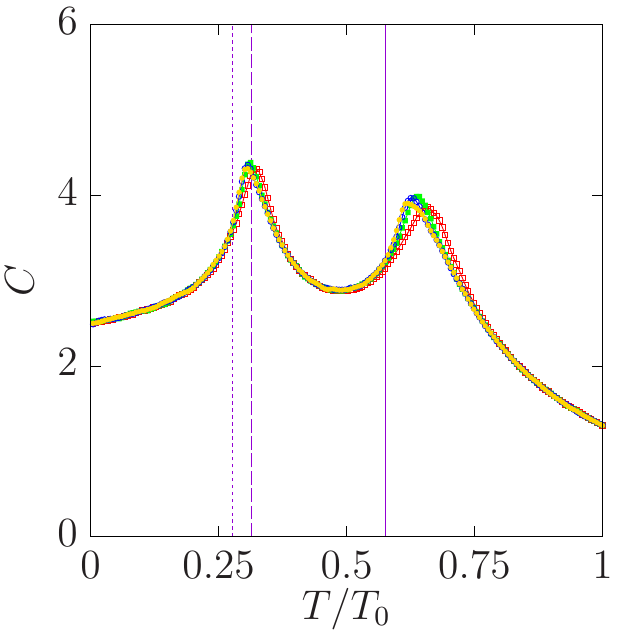}
\end{minipage}
\begin{minipage}{0.492\linewidth}
\centering
(d) \\
\includegraphics[height=0.95\linewidth]{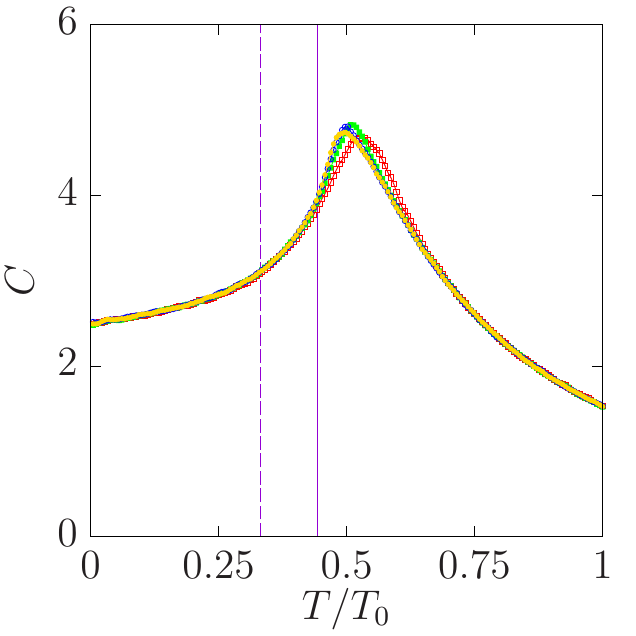}
\end{minipage}
\caption{
\label{fig:specific-heat-A20}
Temperature dependence of the specific heat $C$ for (a) the polar state with $\upsilon = 0^\circ$, (b) the polar state with $\upsilon = 45^\circ$, (c) the broken-axisymmetric state with $\upsilon = 150^\circ$, and (d) the anti-ferromagnetic state with $\upsilon = -45^\circ$.
The violet solid and dashed lines show the mass BKT transition temperature $T_{\rm mass}^{\rm BKT}$ and the spin-singlet crossover temperature $T_{\rm singlet}^{\rm CO}$, respectively.
The violet dash-dot line in panel (c) shows the spin BKT transition temperature $T_{\rm spin}^{\rm BKT}$.
}
\end{figure}
Figure \ref{fig:specific-heat-A20} shows the specific heat $C \equiv (1/L^2) d\langle \mathcal{H} \rangle / d T$, and there is no structure characterizing the emergence of the quasi off-diagonal long-range order of the spin-singlet paring.
This result suggest that the emergence of the quasi off-diagonal long-range order of the spin-singlet paring is not the transition but the crossover.
We here define the spin-singlet crossover temperature $T_{\rm singlet}^{\rm CO}$ at which the spin-singlet paring correlation function satisfies $G_{A_{20}} \propto r^{-1/4}$.
In Figs. \ref{fig:u1-rhosm} (b), \ref{fig:u1u12-rhos} (a), \ref{fig:BA-rhos} (a), and \ref{fig:polar0-rhos} (a) showing the temperature dependencies of the mass superfluid density $\Upsilon_1$, we also show the spin-singlet crossover temperature $T_{\rm singlet}^{\rm CO}$.
As well as the specific heat $C$, the mass superfluid density $\Upsilon_1$ does not drastically change at the spin-singlet crossover temperature $T_{\rm singlet}^{\rm CO}$.
However, it can be clearly seen that the four-times larger universal relation
\begin{align}
\Upsilon_1 = \left(\frac{8 M}{\pi \hbar^2}\right) T_{\rm singlet}^{\rm CO},
\label{eq:universal-relation-spin-singlet}
\end{align}
holds at the spin-singlet crossover temperature $T_{\rm singlet}^{\rm CO}$ for any $\upsilon$.
We note that $T_{\rm singlet}^{\rm CO}$ is not the transition temperature, the jump $\Delta \Upsilon_1$ of the mass superfluid density for the conventional universal relation is replaced by the mass superfluid density $\Upsilon_1$ itself.
From this result, we expect that the quasi off-diagonal long-range order of the spin-singlet paring has some contribution to the mass superfluid density.

\subsection{Additional simulations with other numerical parameters}
\label{subsec:remarks}

\subsubsection{Absence of BKT transition with $\mathbb{Z}_2$ vortices}

\begin{figure}[htb]
\centering
\begin{minipage}{0.492\linewidth}
\centering
(a) \\
\includegraphics[height=0.95\linewidth]{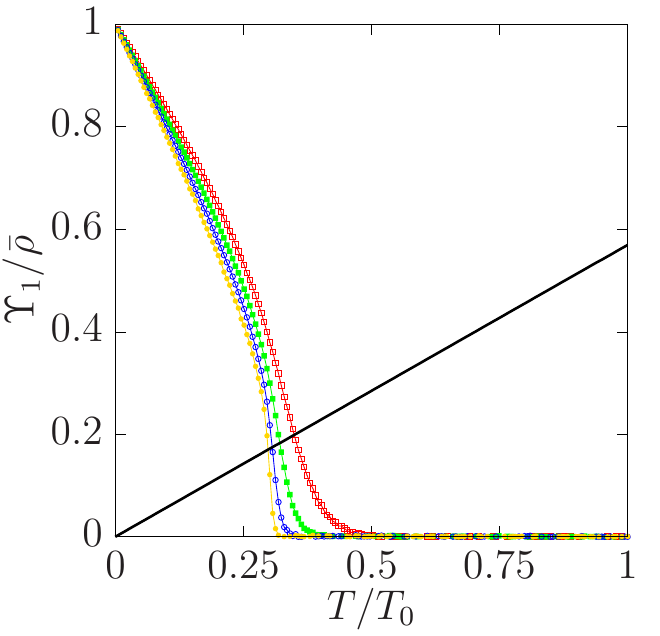}
\end{minipage}
\begin{minipage}{0.492\linewidth}
\centering
(b) \\
\includegraphics[height=0.95\linewidth]{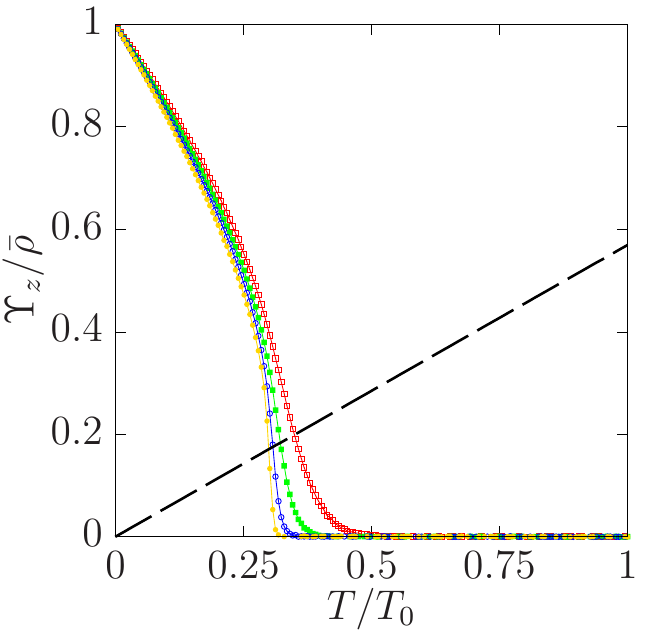}
\end{minipage}
\caption{
\label{fig:Z2-compressible}
Temperature dependence of (a) the mass superfluid density $\Upsilon_1$ and the spin superfluid density $\Upsilon_z$ for the ferromagnetic state with $\tilde{g} = 0.05 g_1$ and $\upsilon = 180^\circ$.
The black solid and dashed lines in panels (a) and (b) show the relation $\Upsilon_1 / T = 2 M / (\pi \hbar^2)$ and $\Upsilon_z / T =  2 M / (\pi \hbar^2)$, respectively.
}
\end{figure}

An important conclusion of our results is that the BKT transition occurs only when the manifold of the ground state has a $\mathrm{S}^1$ part and the vortices are classified by the integer group $\mathbb{Z}$.
Even when $\mathbb{Z}_2$ vortices can exist in the manifold of $\mathbb{R}\mathrm{P}^2$ for the polar state and $\mathbb{R}\mathrm{P}^3$ for the ferromagnetic state without the quadratic Zeeman effect, the BKT transition is apparently absent.
Our results contradict studies for the anti-ferromagnetic Heisenberg model on a two-dimensional triangular lattice \cite{Kawamura-so3} and the simple $\mathrm{SO}(3)$ model \cite{Kikuchi} in which the manifold of the ground state is $\mathbb{R}\mathrm{P}^3$ with $\mathbb{Z}_2$ vortices and some kind of phase transition is predicted.
A possible reason for this contradiction is that the difference between our model for the spinor Bose system and other models \cite{difference-SO3} may be crucial for the existence of the transition.
We note that the absence of the BKT transition is apparent even with the smaller spin-dependent inter-particle interaction $\tilde{g} = 0.05 g_0$ as shown in Fig. \ref{fig:Z2-compressible}, where the state can more easily escape from the $\mathrm{SO}(3)$ manifold under the temperature fluctuation. 

\subsubsection{Universal relation of the superfluid density}

Another conclusion is the universal relation of the superfluid density.
In particular, a universal relation becomes two-times and three-times larger for the anti-ferromagnetic state ($q < 0$) and the polar state ($q = 0$) without the quadratic Zeeman effect, respectively than the polar state with the positive quadratic Zeeman effect [see Eqs. \eqref{eq:mass-universal-relation}, \eqref{eq:large-universal-relation-mass}, and \eqref{eq:large0-universal-relation-mass}].
This result becomes rather unclear when both the quadratic Zeeman effect $q$ and the spin-dependent inter-particle interaction $g_1$ are small.
\begin{figure}[htb]
\centering
\begin{minipage}{0.492\linewidth}
\centering
(a) \\
\includegraphics[height=0.95\linewidth]{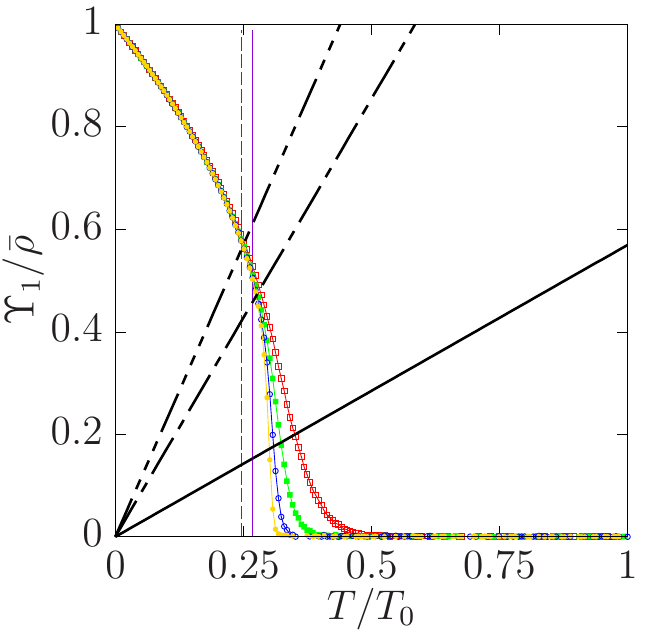}
\end{minipage}
\begin{minipage}{0.492\linewidth}
\centering
(b) \\
\includegraphics[height=0.95\linewidth]{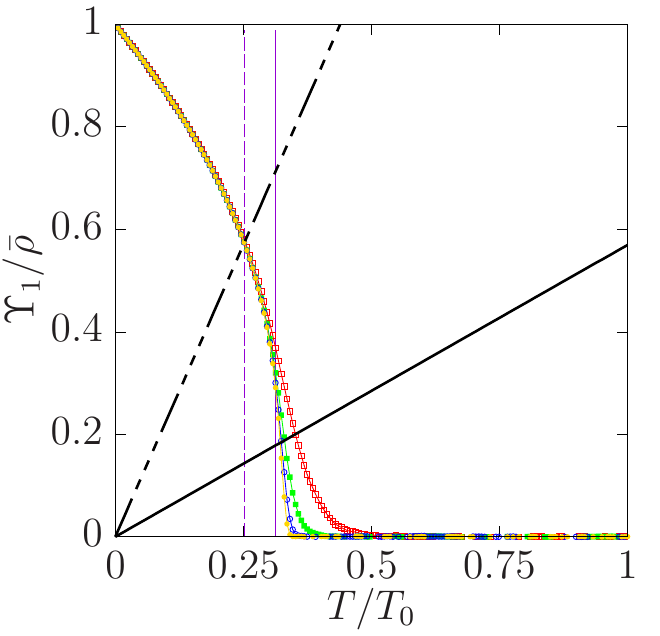}
\end{minipage} \\[10pt]
\begin{minipage}{0.492\linewidth}
\centering
(c) \\
\includegraphics[height=0.95\linewidth]{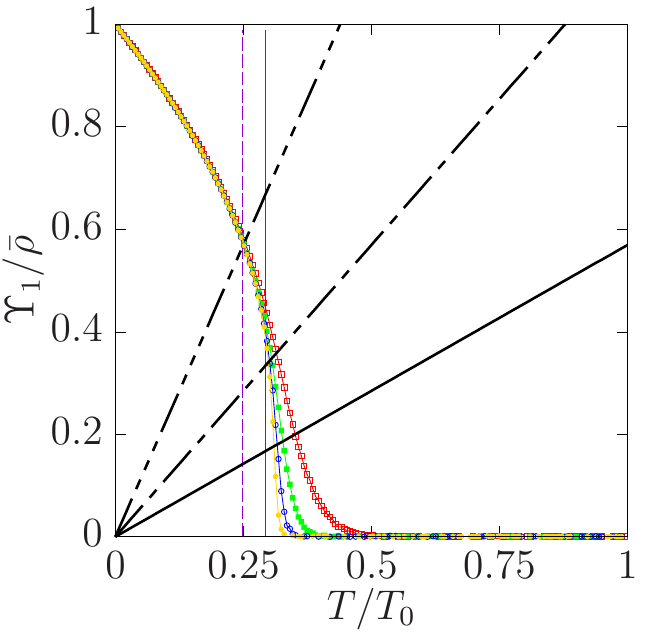}
\end{minipage}
\begin{minipage}{0.492\linewidth}
\centering
(d) \\
\includegraphics[height=0.95\linewidth]{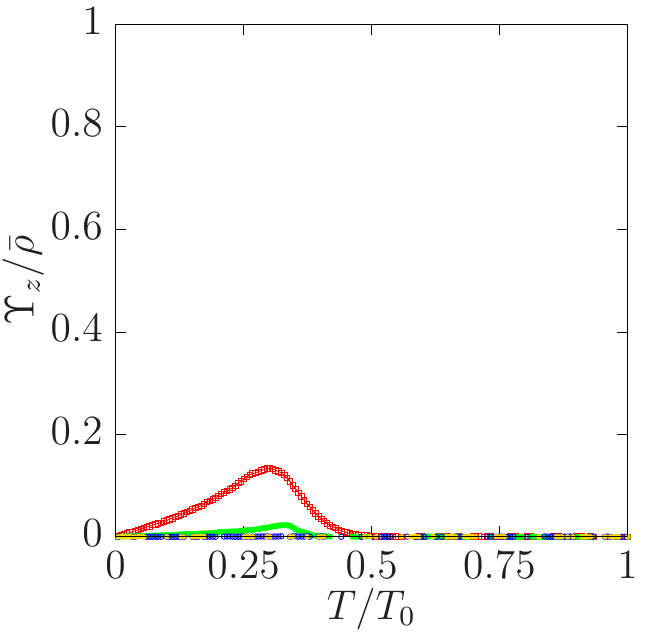}
\end{minipage}
\caption{
\label{fig:compressible}
Temperature dependence of the mass superfluid density $\Upsilon_1$ for (a) the polar state with $\upsilon = 0^\circ$, (b) the polar state with $\upsilon = 1^\circ$, and (c) the anti-ferromagnetic state with $\upsilon = -1^\circ$, and (d) the spin superfluid density $\Upsilon_z$ for the polar state with $\upsilon = 1^\circ$.
For all panels, we use $\tilde{g} = 0.05 g_1$.
The violet solid and dashed lines in panels (a)-(c) show the mass BKT transition temperature $T_{\rm mass}^{\rm BKT}$ and the spin-singlet crossover temperature $T_{\rm singlet}^{\rm CO}$, respectively.
The black solid lines in panels (a)-(c) show the relation $\Upsilon_1 / T = 2 M / (\pi \hbar^2)$.
The black dash-dot lines in panels (a) and (c) show the relation $\Upsilon_1 / T =  6 M / (\pi \hbar^2)$ and $\Upsilon_1 / T =  4 M / (\pi \hbar^2)$, respectively.
The black three-dot-chain lines in panels (a)-(c) show the relation $\Upsilon_1 / T =  8 M / (\pi \hbar^2)$.
}
\end{figure}
Figures \ref{fig:compressible} (a)-(c) shows the mass superfluid density for $\tilde{g} = 0.05 g_1$ and $\upsilon = 0^\circ$ [in panel (a)], $\upsilon = 1^\circ$ [in panel (b)], and $\upsilon = -1^\circ$ [in panel (c)].
Whereas the three-times larger universal relation is still apparent in panel (a), universal relations in panels (b) and (c) look larger than the standard and two-times larger ones respectively.
Although these results suggest the crossover behavior of the universal relation at the phase boundary with $q = 0$, more detailed numerical analyses are needed to fix this problem.

\subsubsection{Absence of two-step transition in polar state with $q > 0$}

Other numerical work on the polar state \cite{James} has reported two-step phase transitions at different temperatures when the quadratic Zeeman effect is very weak.
A nontrivial two-step phase transition in a two-dimensional system has been reported for a many-particle system \cite{Halperin,Young,Zahn}, the $n$-clock model with $n \geq 3$ \cite{Kumano}, and the modified $XY$-model \cite{Carpenter}.
Although the two-step phase transitions of these models are triggered by the gradient term of the Hamiltonian, the authors of the above work on the polar state have shown that the two-step transition is caused by the quadratic Zeeman effect as an external symmetry-breaking field.
As a feature of the two-step phase transition, the authors report a double peak in the temperature dependence of the specific heat and intersections of the binder cumulant at two different temperatures.
A four-times larger universal relation of the mass superfluid density has also been reported.
These results indicate that the transition is quite abnormal compared with the standard BKT transition, as shown in Fig. \ref{fig:u1-rhosm} (b).
However, we have not been able to confirm their results even when $\tilde{g} = 0.05$ and $\upsilon = 1^\circ$, as shown in Figs. \ref{fig:compressible} (b) and (d), which shows that there is a simple BKT transition for the mass superfluid density, and the spin superfluid density is absent at all temperatures as for the case of $\tilde{g} = 0.5$ and $\upsilon = 45^\circ$ [see Figs. \ref{fig:u1-rhosm} (b) and \ref{fig:u1-rhoss}].
\begin{figure}[htb]
\centering
\begin{minipage}{0.492\linewidth}
\centering
  (a) \\
\includegraphics[height=0.95\linewidth]{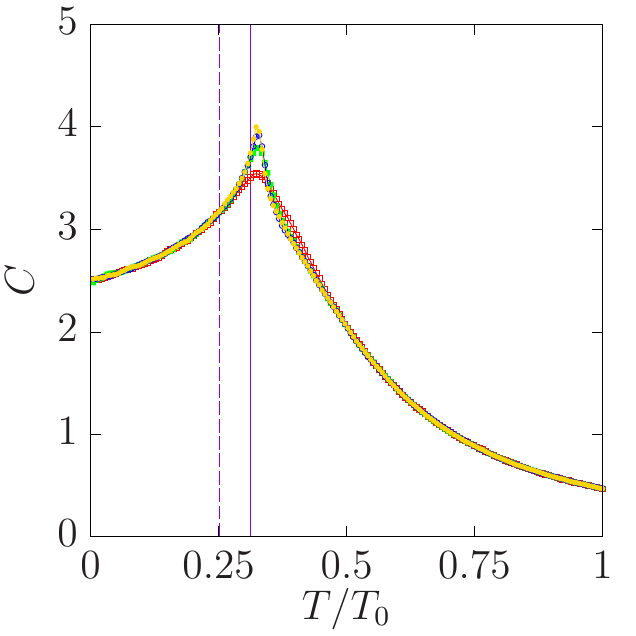}
\end{minipage}
\begin{minipage}{0.492\linewidth}
\centering
 (b) \\
\includegraphics[height=0.95\linewidth]{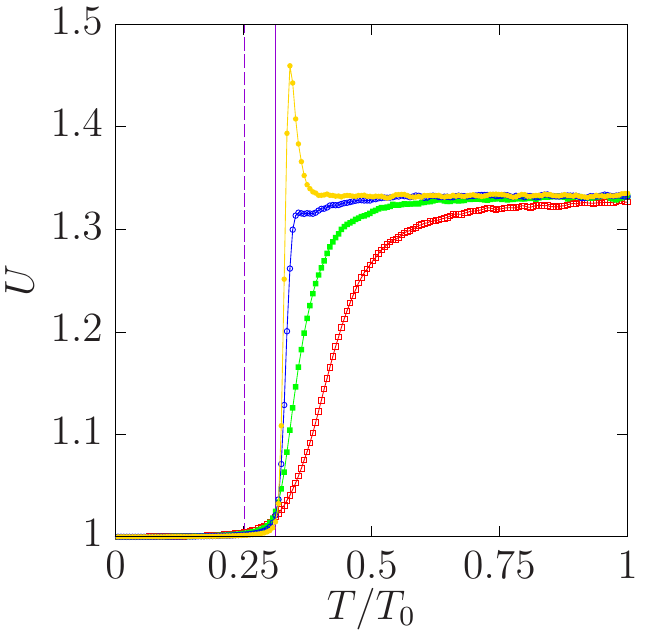}
\end{minipage}
\caption{
\label{fig:Um-5circ}
(a) Temperature dependence of the specific heat $C$ and (b) the binder cumulant $U$ for the polar state with $\tilde{g} = 0.05$ and $\upsilon = 1^\circ$.
The violet solid and dashed lines in both panels show the mass BKT transition temperature $T^{\rm BKT}_{\rm mass} \approx 0.31 T_0$ and the spin-singlet crossover temperature $T_{\rm singlet}^{\rm CO} \approx 0.25 T_0$, respectively.
}
\end{figure}
In Fig. \ref{fig:Um-5circ}, we show the specific heat $C$ and the binder cumulant $U$, defined as
\begin{align}
U \equiv \frac{\displaystyle \langle \Psi^4 \rangle}{\displaystyle \langle \Psi^2 \rangle^2}, \quad
\Psi^2 \equiv \frac{1}{L^4} \sum_{m=-1}^1 \left| \int d^2x\: \psi_m \right|^2.
\end{align}
The specific heat has a single peak just above the mass BKT transition temperature $T^{\rm BKT}_{\rm mass}$ and the binder cumulant intersects only once near the mass BKT transition temperature, which supports the simple BKT transition.
A possible reason for the discrepancy is that the system is still not equilibrated in the work.
In the region around $\upsilon = 0^\circ$, there is a change of the ground-state manifold between $S^1$ and $(S^1 \times S^2)/\mathbb{Z}_2$, and we expect a rapid increase of the relaxation time.
It has been theoretically predicted that two-step transitions due to two different topological defects are possible in a non-equilibrated system \cite{Sasa}, and the authors of the work may have observed two-step nonequilibrium transitions.
The absence of a two-step transition has also been reported for two-component Bose mixtures with inter-component Josephson coupling \cite{Kobayashi}.

\section{Summary and Discussion}
\label{sec:summary}

We have numerically studied BKT transitions for mass and spin superfluidies of a spin-1 spinor Bose system under the quadratic Zeeman effect.
The qualitative properties of these transitions strongly depend on the sign and strength of the spin-dependent coupling constant and the quadratic Zeeman effect, and are determined by the manifold of the ground state.

\begin{figure}[htb]
\centering
\includegraphics[height=0.6\linewidth]{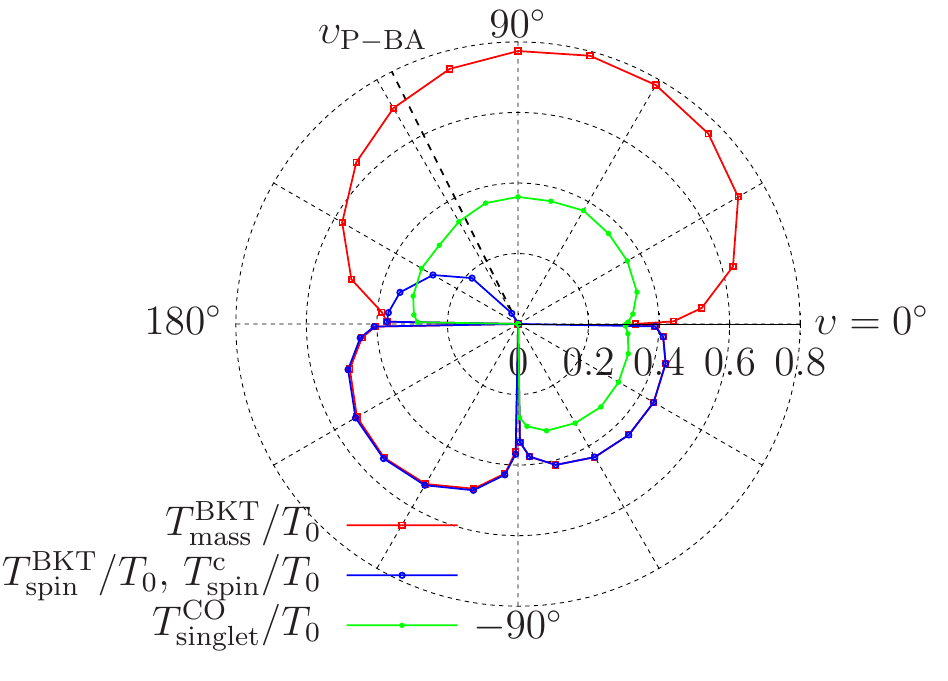}
\caption{
\label{fig:TBKT}
Dependence of the mass and spin BKT transition temperatures $T^{\rm BKT}_{\rm mass}$ (red) and $T^{\rm BKT}_{\rm spin}$ (blue), and the spin-singlet crossover temperature $T_{\rm singlet}^{\rm CO}$ (green) on $\upsilon$.
For the ferromagnetic state with $-180^\circ < \upsilon < - 90^\circ$, the spin thermodynamic transition temperature $T^{\rm c}_{\rm spin}$ is shown instead of the spin BKT transition temperature $T^{\rm BKT}_{\rm spin}$.
  }
\end{figure}

The numerically obtained mass and spin BKT transition temperatures $T^{\rm BKT}_{\rm mass}$ and $T^{\rm BKT}_{\rm spin}$ are summarized in Fig. \ref{fig:TBKT}.

For the polar state with $\upsilon = 0^\circ$, only the mass BKT transition without spin superfluid density is observed.
The mass superfluid density $\Upsilon_1$ shows a three-times larger universal relation in Eq. \eqref{eq:large0-universal-relation-mass}.

For the polar state with $0^\circ < \upsilon \leq \upsilon_{\rm P-BA}$, again only the mass BKT transition is observed, but the mass superfluid density $\Upsilon_1$ follows the standard universal relation in Eq. \eqref{eq:mass-universal-relation}.

The broken-axisymmetric state with $\upsilon_{\rm P-BA} < \upsilon < 180^\circ$ exhibits independent mass and spin BKT transitions with different BKT transition temperatures, $T^{\rm BKT}_{\rm mass}$ and $T^{\rm BKT}_{\rm spin}$.
The mass BKT transition temperature $T_{\rm mass}^{\rm BKT}$ continuously changes at the phase boundary with $\upsilon = \upsilon_{\rm P-BA}$ and suddenly vanishes at $\upsilon = 180^\circ$.
The spin BKT transition temperature $T_{\rm spin}^{\rm BKT}$ also continuously grows at $\upsilon = \upsilon_{\rm P-BA}$ and suddenly vanishes at $\upsilon = 180^\circ$, and always lower than the mass BKT transition temperature $T_{\rm mass}^{\rm BKT}$.
Both the mass and spin superfluid densities $\Upsilon_1$ and $\Upsilon_z$ satisfy the standard universal relations in Eqs. \eqref{eq:mass-universal-relation} and \eqref{eq:spin-universal-relation}.

For the ferromagnetic state with $\upsilon = 180^\circ$, both transitions are absent.

For the ferromagnetic state with $-180^\circ < \upsilon < -90^\circ$, the mass BKT transition and the spin thermodynamic transition are observed.
The two transition temperatures $T^{\rm BKT}_{\rm mass}$ and $T^{\rm c}_{\rm spin}$ for this state are very close to each other for the whole range of $\upsilon$, and we do not observe any evidence of the difference for the present numerical accuracy.
Both the mass and spin superfluid densities $\Upsilon_1$ and $\Upsilon_z$ satisfy the standard universal relations in Eqs. \eqref{eq:mass-universal-relation} and \eqref{eq:spin-universal-relation}.

For the $\mathrm{SU}(2)$-symmetric state with $\upsilon = -90^\circ$, both transitions are absent.

For the anti-ferromagnetic state with $-90^\circ < \upsilon < 0^\circ$, the mass and spin BKT transitions are observed with the same transition temperature.
Both the mass and spin superfluid densities $\Upsilon_1$ and $\Upsilon_z$ satisfy two-times larger universal relations in Eqs. \eqref{eq:large-universal-relation-mass} and \eqref{eq:large-universal-relation-spin}.

An important conclusion is the difference between the two transition temperatures $T^{\rm BKT}_{\rm mass}$ and $T^{\rm BKT}_{\rm spin}$ (or $T^{\rm c}_{\rm spin}$).
For the case of a positive quadratic Zeeman effect with $0^\circ < \upsilon < 180^\circ$, the two transition temperatures are different from each other, whereas they are the same for a negative quadratic Zeeman effect with $-180^\circ < \upsilon < 0^\circ$ except for $\upsilon = -90^\circ$.
The origin of the similarity between the transition temperatures is different for the anti-ferromagnetic state with $-90^\circ < \upsilon < 0^\circ$ and the ferromagnetic state with $-180^\circ < \upsilon < -90^\circ$.
For the anti-ferromagnetic state, the two transition temperatures are the same simply because of the equivalence between the ground state \eqref{eq:polar--} and the two-component Bose system with an equal mass and density.
For the ferromagnetic state with $-180^\circ < \upsilon < -90^\circ$, the situation is more complicated.
The manifold $\mathrm{O}(2) \cong \mathbb{Z}_2 \ltimes \mathrm{SO}(2)$ of the ground state can be separated into a continuous $\mathrm{SO}(2) \simeq S^1$ part for the global phase and a discrete $\mathbb{Z}_2$ part for the spin angle.
However, the two partial manifolds $\mathrm{SO}(2)$ and $\mathbb{Z}_2$ are not simply connected with the direct product, but rather are connected with the non-commutative semi-direct product \cite{direct-product}.
The fact that the two transition temperatures are the same (or very close) may arise from the connection of the partial manifolds with the semi-direct product.
In a related study, an analysis of the anti-ferromagnetic $XY$-model having a $\mathrm{O}(2)$ ground-state manifold \cite{Miyashita,Kawamura-o2,Obuchi} found that the two transition temperatures for the spin and chiral ordering are very close but are different, and the universality class for the spin ordering is apparently different from that for the standard $XY$-model.
Another related topic is the analysis of the Heisenberg model for a distorted triangular lattice \cite{Tamura} in which the manifold of the ground state is a coupling of a continuous $\mathbb{R}\mathrm{P}^3$ part and a discrete $\mathbb{Z}_2$ part.
For this model, it has been reported that the thermodynamic phase transition from the $\mathbb{Z}_2$ part with the Ising universality class and the dissociation of the $\mathbb{Z}_2$ vortices from the $\mathbb{R}\mathrm{P}^3$ part occur at the same temperature.
We expect that these results are closely related with our result for $T^{\rm BKT}_{\rm mass} \fallingdotseq T^{\rm c}_{\rm spin}$.
A spin-triplet superconductor is another related system \cite{Babaev}.
The macroscopic Hamiltonian is very similar to that for the spin-1 spinor Bose gas except for the local gauge invariance of the Hamiltonian which plays an important role to determine properties of the BKT transition.
While our results show the mass BKT transition temperature is always larger than the spin BKT transition as $T^{\rm BKT}_{\rm mass} \geq T^{\rm BKT}_{\rm spin}$ for all $\upsilon$, it has been reported that the local gauge invariance can induce the situation of $T^{\rm BKT}_{\rm mass} < T^{\rm BKT}_{\rm spin}$, i.e., spin superfluidity without superconductivity.

Another conclusion concerns the height of the universal relation of the superfluid density.
Here, we suggest that the height of the universal jump is determined by how many kinds of vortices contribute to the BKT transition.
For the anti-ferromagnetic state, two different half-quantized vortices shown in Eqs. \eqref{eq:polar-half-vortex-1} and \eqref{eq:polar-half-vortex-2} contribute to the BKT transition at the BKT transition temperature $T_{\rm mass}^{\rm BKT} = T_{\rm spin}^{\rm BKT}$, inducing the two-times larger universal relation.
For the polar state with the positive quadratic Zeeman effect, the only integer vortex shown in Eq. \eqref{eq:polar-+-vortex} contributes to the BKT transition and the standard universal relation holds.
For the polar state without the quadratic Zeeman effect, all of the half-quantized vortices appeared in the anti-ferromagnetic state and the integer vortex appeared in the polar state with the positive quadratic Zeeman effect contribute to the BKT transition, inducing the three-times larger universal relation.
For the broken-axisymmetric state, the phase vortex shown in Eq. \eqref{eq:BA-phase-vortex} and the spin vortex shown in Eq. \eqref{eq:BA-spin-vortex} separately contribute to the mass BKT transition and the spin BKT transition, respectively, inducing the standard universal relations at the mass and spin BKT transition temperatures $T_{\rm mass}^{\rm BKT}$ and $T_{\rm spin}^{\rm BKT}$ independently.
For the ferromagnetic state with the negative quadratic Zeeman effect, under the $\mathbb{Z}_2$ symmetry breaking, either integer vortex shown in Eq. \eqref{eq:Ferro--vortex-1} or that shown in Eq. \eqref{eq:Ferro--vortex-2} contributes to the mass BKT transition, inducing the standard universal relation.
We cannot have any clear answer to why the spin BKT transition also satisfy the standard universal relation (we cannot omit the possibility in which the spin superfluid density weakly deviates from the standard universal relation).

We have also found another four-times larger universal relation \eqref{eq:universal-relation-spin-singlet} at the spin-singlet crossover temperature $T_{\rm singlet}^{\rm CO}$.
In the previous work \cite{Mukerjee}, it has been predicted that the quasi off-diagonal long-range order of the spin-singlet paring emerges at the BKT transition temperature and gives the four-times larger universal relation for the mass superfluid density for the polar state without the quadratic Zeeman effect.
Although author's claim about the relationship between the quasi off-diagonal long-range order of the spin-singlet paring and the four-times larger universal relation is correct, we problematize that authors have treated the spin-singlet crossover temperature $T_{\rm singlet}^{\rm CO}$ as the mass BKT transition temperature $T_{\rm mass}^{\rm mass}$ due to the fact that they are close to each other for the polar state at $\upsilon = 0^\circ$.

There is one more important open issue to be further analyzed in detail.
Figure \ref{fig:TBKT} shows that the BKT transition temperatures $T_{\rm mass}^{\rm BKT}$ and $T_{\rm spin}^{\rm BKT}$ rapidly changes at the phase boundaries with $\upsilon = 0^\circ$ (between anti-ferromagnetic and polar states), $\upsilon = 180^\circ$ (between broken-axisymmetric and ferromagnetic states), and $\upsilon = -90^\circ$ (between ferromagnetic and anti-ferromagnetic states).
In particular, the mass (spin) BKT transition suddenly vanish at $\upsilon = -90^\circ$ and $\upsilon = 180^\circ$ ($\upsilon = 0^\circ$, $\upsilon = -90^\circ$, and $\upsilon = 180^\circ$).
This is because the manifold $\mathcal{M}$ of the ground state discontinuously changes at these phase boundaries.
On the other hand, the BKT transition temperatures continuously changes at the phase boundary with $\upsilon = \upsilon_{\rm P-BA}$ between the polar and broken-axisymmetric states because the change of the manifold $\mathcal{M}$ from $\mathrm{S}^1$ to $\mathrm{S}^1 \times \mathrm{S}^1$ is continuous.
In the present work, we cannot determine whether the rapid changes of the transition temperatures at $\upsilon = 0^\circ$, $180^\circ$, $-90^\circ$ are still continuous as a crossover, like the continuous clock model \cite{Jose}, or discontinuous as ``a transition of the BKT transition''.
Compared to the mass and spin BKT transition temperatures $T_{\rm mass}^{\rm BKT}$ and $T_{\rm spin}^{\rm BKT}$,the variation of the spin-singlet crossover temperature $T_{\rm singlet}^{\rm CO}$ is small except for the boundaries at $\upsilon = 180^\circ$ (between broken-axisymmetric and ferromagnetic states) and $\upsilon = -90^\circ$ (between ferromagnetic and anti-ferromagnetic state).%, and the same problem arises at these boundaries.

Future work will concentrate on spin-2 spinor Bose gases.
In a spin-2 system, we can expect richer vortices, such as vortices having third and fourth-fractionalized circulation, non-Abelian vortices, and $\mathbb{Z}_4$ vortices \cite{Kawaguchi}.
The existence of the BKT transition and the relationship between the fractional circulation and the universal relation of the superfluid density will be better clarified in these upcoming studies.
The effect of quantum fluctuations will also be the subject of future studies.
In this paper, we neglect quantum fluctuations because BKT transitions occur at finite temperatures.
However, for states in which the BKT transition temperature vanishes with $T^{\rm BKT}_{\rm mass,spin} = 0$, such as states for $\upsilon = 0^\circ$, $180^\circ$, and $-90^\circ$, the disappearance of superfluidity may be corrected by quantum fluctuations.
A related work has studied one-dimensional spinor Bose gas at zero temperature \cite{Konig}.
Quantum fluctuations are controlled by the spin-dependent inter-particle interaction $g_2$, and order-disorder quantum phase transitions occur as a function of $g_2$ for nonzero quadratic Zeeman effect $q \neq 0$.
Two ordered phases with positive and negative $q$ are continuously connected by a Luttinger liquid with $q=0$, and the transition from Luttinger liquid to disordered phase occurs as the BKT transition, which suggests the importance of quantum fluctuations near $q = 0$.
Another related work has studied two-dimensional spinor Bose gas in an optical lattice \cite{Podolsky}.
In this system, the optical lattice amplifies the strength of quantum fluctuation, and it has been reported that the equivalence between mass and spin superfluid densities in the anti-ferromagnetic state is broken due to quantum fluctuation.
%Furthermore, Bose-Einstein condensation of singlet-paired Bose particles has been predicted for $\upsilon = 0^\circ$ in three-dimensional systems \cite{Kawaguchi,Yukalov,Koashi} due to quantum fluctuations, and we may expect competition between BKT transitions of spinor Bose particles with vortices having half-quantized circulation and singlet-paired Bose particles with vortices having integer circulation.

%quantum fluctuation, spin-2 system

\section*{Acknowledgement}

We would like to thank Shu Tanaka, Koji Hukushima, Blair Blakie, Yuki Kawaguchi, and Shin-ichi Sasa for the helpful suggestions and comments.
%We also thank the anonymous referee for the helpful and constructive comments.
This work is supported by the Japan Society for the Promotion of Science (JSPS) Grant-in-Aid for Scientific Research (KAKENHI Grant No. 16H03984).
The work is also supported in part by a Grant-in-Aid for Scientific Research (KAKENHI Grant Numbers 26870295, 16KT0127).


\begin{thebibliography}{100}
\bibitem{difference-SO3}
A simple discrete $\mathbb{R}\mathrm{P}^3$ model is given by
\begin{align}
\mathcal{H} = - \sum_{\langle i,j \rangle} (\Vec{X}_i \cdot \Vec{X}_j )^2,
\end{align}
where $\Vec{X}_i$ is the four-dimensional unit vector at the two-dimensional lattice point $i$, and $\langle i,j \rangle$ is a pair of nearest lattice points $i$ and $j$.
On the other hand, the discretized limit of the Hamiltonian \eqref{eq:Hamiltonian} for the ferromagnetic ground state \eqref{eq:Ferro-0} without the quadratic Zeeman effect is given by
\begin{align}
  \begin{split}
    & \mathcal{H} = - \sum_{\langle i,j \rangle} \left\{ (\Vec{X}_i \cdot \Vec{X}_j )^2 - (\Vec{X}_i \cdot R \Vec{X}_j)^2 \right\}, \\
    & R = \begin{pmatrix} 0 & 1 \\ -1 & 0 \end{pmatrix} \oplus \begin{pmatrix} 0 & 1 \\ -1 & 0 \end{pmatrix},
  \end{split}
\end{align}
where
\begin{align}
\Vec{X} = \begin{pmatrix}
\cos(\alpha + \gamma)/2 \cos\beta/2 \\
\sin(\alpha + \gamma)/2 \cos\beta/2 \\
\cos(\alpha - \gamma)/2 \sin\beta/2 \\
\sin(\alpha - \gamma)/2 \sin\beta/2
\end{pmatrix}.
\end{align}
\bibitem{direct-product}
A simple model giving the ground-state manifold of the commutative direct product $\mathbb{Z}_2 \times S^1$ is given by
\begin{align}
\mathcal{H} = \int d^2x\: \left\{ |\nabla \Vec{X}|^2 + \frac{g_0}{2} (X_1^2 + X_2^2)^2 + \frac{g_1}{2} X_3^4 \right\},
\end{align}
with the three-dimensional real vector $\Vec{X} = (X_1, X_2, X_3)$.
The discretized limit of this Hamiltonian is given by
\begin{align}
\mathcal{H} = - \sum_{\langle i,j \rangle} \left\{ \cos(\alpha_i - \alpha_j) + \delta_{\phi_i,\phi_j} \right\},
\label{eq:direct-product}
\end{align}
where $(X_1, X_2)$ is normalized as $(X_1, X_2) = (\cos\alpha,\sin\alpha)$ and $X_3$ is also normalized as $X_3 = \phi = \pm 1$.
On the other hand, the discretized limit of the Hamiltonian \eqref{eq:Hamiltonian} for the ferromagnetic ground state \eqref{eq:Ferro--} is given by
\begin{align}
\mathcal{H} = - \sum_{\langle i,j \rangle} \cos(\alpha_i - \alpha_j) \delta_{\phi_i,\phi_j}.
\label{eq:semi-direct-product}
\end{align}
We can easily see that the continuous and discrete degrees of freedom $\alpha$ and $\phi$ are independent of each other in Eq. \eqref{eq:direct-product} and coupled to each other in Eq. \eqref{eq:semi-direct-product}.
\end{thebibliography}
\end{document}